\documentclass[12pt]{article}
\usepackage{epsfig}
\usepackage[dvips]{color}
\newcommand{\bla}{\textcolor{bla}}
\newcommand{\rot}{\textcolor{rot}}
\definecolor{bla}{rgb}{0,0,1}
\definecolor{rot}{rgb}{1,0,0}
\newcommand{\pp}{\phantom{-}}
\newcommand{\zz}{\phantom{2}}

\newcommand{\Li}{\mbox{Li}_2}
\newcommand{\Ti}{\mbox{Li}_3}

\newcommand{\1}{1 \hspace{-1.05mm} {\rm l}}
\newcommand{\Lbf}{\Large \bf}
\newcommand{\pbm}{\protect \boldmath}
\newcommand{\slp}{p \hspace{-2mm} /}
\newcommand{\sls}{s \hspace{-2mm} /}
\newcommand{\slk}{k \hspace{-2mm} /}

\newenvironment{Eqnarray}{\arraycolsep 0.14em \begin{eqnarray}}{\end{eqnarray}}
\newenvironment{Eqnarray*}{\arraycolsep 0.14em \begin{eqnarray*}}
{\end{eqnarray*}}
\renewcommand{\theequation}{\mbox{\arabic{equation}}}
\newcounter{saveeqn}

\begin{document}
\thispagestyle{empty}
\begin{flushright}
        MZ--TH/02--03\\
        hep--ph/0203048\\
        January 23, 2003\\
\end{flushright}
\vspace{5mm}
\begin{center}
 {\Lbf Leptonic \pbm $ \mu $- and \pbm $ \tau $-decays: mass effects,}\\[3mm]
 {\Lbf polarization effects and \pbm $ O(\alpha) $ radiative corrections}\\[7mm]
 {\large M.~Fischer, S.~Groote, J.G.~K\"orner and M.C.~Mauser}\\[7mm]
       Institut f\"ur Physik, Johannes Gutenberg--Universit\"at\\[2mm]
       Staudinger Weg 7, D--55099 Mainz, Germany\\[25mm]
\end{center}


\begin{abstract}
\noindent
 We calculate the radiative corrections to the unpolarized and the
 four polarized spectrum and rate functions in the leptonic decay of a
 polarized $ \mu $ into a polarized electron.
 The new feature of our calculation is that we keep the mass of the
 final state electron finite which is mandatory if one wants to investigate
 the threshold region of the decay.
 Analytical results are given for the energy spectrum and the
 polar angle distribution of the final state electron whose
 longitudinal and transverse polarization is calculated.
 We also provide analytical results on the integrated spectrum functions.
 We analyze the $ m_e \rightarrow 0 $ limit of our general results and
 investigate the quality of the $ m_e \rightarrow 0 $ approximation.
 In the $ m_e \rightarrow 0 $ case we discuss in some detail the
 role of the $ O(\alpha) $ anomalous helicity flip contribution of the
 final electron which survives the $ m_e \rightarrow 0 $ limit.
 The results presented in this paper also apply to the leptonic decays of
 polarized $ \tau $--leptons for which we provide numerical results. 
\end{abstract}

\newpage


\section{\bf Introduction}

 While calculating the radiative $ O(\alpha_s) $ QCD corrections
 to polarization effects in semileptonic decays of heavy quarks,
 where the full quark mass dependence was retained \cite{fgkm00,fgkm01},
 we came to realize that our results could also be gainfully employed
 in the corresponding $ O(\alpha) $ QED corrections to the weak leptonic
 decays of the $ \mu $-- and $ \tau $--leptons\footnote{
 The three leptonic $ \mu $-- and $ \tau $--decays are treated within the
 Standard Model, i.e. they are all governed by the same $ (V - A) $
 coupling structure and coupling strength $ G_F $.}.
 In most of the previous radiative correction calculations the
 mass of the charged lepton daughter $ l' $ has been neglected except for
 anomalous contributions from the collinear region which survive the
 $ m_l' \rightarrow 0 $ limit \cite{lee64,spinflip,falk94} and the
 logarithmic terms $ \sim (\ln m_l') $ which are needed to regularize the
 collinear divergencies that appear in the loop and tree graph
 (``internal bremsstrahlung'') contributions.
 These logarithmic terms partially cancel in the spectrum and
 completely cancel in the rate when the loop and tree graph contributions
 are added.

 From general considerations it follows that the unpolarized and three
 of the polarized spectrum functions contain only even powers of the
 mass ratio $ m_l' / m_l $.
 Considering the fact that
  $ (m_{e} / m_{\mu})^2    = 2.34 \cdot 10^{-5} $, 
  $ (m_{\mu} / m_{\tau})^2 = 3.54 \cdot 10^{-3} $ and
  $ (m_{e} / m_{\tau})^2   = 8.27 \cdot 10^{-8} $
 the zero mass approximation should be a good approximation
 for most of the energy spectrum of the daughter leptons except
 for the region close (or very close) to the soft endpoint of
 the spectrum (also referred to as the threshold region) where
 finite mass effects have to be retained.
 Contrary to this the transverse polarization of the
 daughter lepton is proportional to the linear mass ratio $ m_l' / m_l $. 
 Also, when integrating the spectrum functions, the linear mass
 ratio enters in all four polarized rate functions.
 Finite mass corrections may thus play an important role at least for
 $ (\tau \rightarrow \mu) $--decays where the linear mass ratio
 $ m_{\mu} / m_{\tau} = 5.95 \cdot 10^{-2} $ is not very small.
 An improved analysis of $ \tau $--decays is of quite some topical
 interest since large samples of $ \tau $--leptons are currently
 being produced at the existing two $ B $ factories in Japan and
 in the USA, and are expected to be produced at future $ \tau $--charm
 factories to be set up in Ithaca and Beijing. 
 As the data becomes more precise, the predictions of the SM
 including also radiative correction effects will be tested at
 an ever--rising level of precision.
 It is then evident that the inclusion of final lepton
 mass effects must play an important role in particular
 in the threshold region.
 
 We determine the radiative corrections to the daughter's lepton energy
 spectrum and its longitudinal and transverse polarization keeping the
 dependence on the polarization of the parent lepton.
 This generalizes the calculation of \cite{fs74} in which the zero mass
 approximation was used.
 Our calculation extends the calculation of \cite{arbuzov01}, which
 also includes finite mass effects, in that we include the longitudinal
 and transverse polarization of the daughter lepton.

 The paper is structured as follows.
 In Sec.~2 we introduce our notation and write down
 the general structure of the spin--dependent rate.
 Sec.~3 contains our Born term results.
 $ W $--propagator effects are taken into account in Sec.~4.
 In Sec.~5 we present our analytical and numerical results on the
 $ O(\alpha) $ radiative corrections.
 In Sec.~6 we consider the $ m_l' \rightarrow 0 $ limit
 of the $ m_l' \ne 0 $ results presented in Sec.~5 and
 discuss in some detail the origin of the anomalous
 helicity flip contribution.
 Sec.~7 contains our summary and conclusions.
 In two Appendices we collect some technical material on
 trilog functions and Fierz identities relevant to our calculation.


\section{General structure of spin-dependent rate}

 To make life simple we shall in the following always refer to the specific
 case $ \mu^{-} \rightarrow e^{-} + \bar{\nu}_{e} + \nu_{\mu} $ instead of
 referring to the generic case involving also leptonic $ \tau $--decays when
 writing down analytical results.
 Of course, when discussing numerical results,
 the two leptonic $ \tau $--decay channels
 $ \tau^{-} \rightarrow \mu^{-} + \bar{\nu}_{\mu} + \nu_{\tau} $ and 
 $ \tau^{-} \rightarrow e^{-} + \bar{\nu}_{e} + \nu_{\tau} $ are also included.

 Since the subject of $ \mu^{-} $--decays is extensively covered in the relevant
 textbooks (see e.g. Refs.~\cite{kaellen64,cb83,scheck83}) and review articles
 (see e.g. Refs.~\cite{lw66,kaellen68,scheck78}) we can afford to be very brief
 in describing the formalism. 

 From helicity counting one knows that there are altogether
 five spin--dependent structure functions and one spin--independent
 structure function describing the leptonic decay of a polarized
 muon into a polarized electron.
 We thus define a spin--dependent differential rate in terms of
 six invariant structure functions $ A_i $.
 In the rest system of the $ \mu^{-} $ the decay distribution reads 

 \begin{Eqnarray} 
   \label{invrate}
   \frac{d \Gamma}{dx \, d \! \cos \theta_P} & = &
   \beta x \, \Gamma_0 \, \bigg( A_1 +
   \frac{1}{m_{\mu}} A_2 (p_{e} \!\cdot\! s_{\mu}) +
   \frac{1}{m_{\mu}} A_3 (p_{\mu} \!\cdot\! s_{e}) +
   \frac{1}{m_{\mu}^2} A_4 (p_e \!\cdot\! s_{\mu}) \,
    (p_{\mu} \!\cdot\! s_{e}) \nonumber \\[2mm] & + &
    A_5 (s_{\mu} \!\cdot\! s_{e}) +
   \frac{1}{m_{\mu}^2} A_6 \,
   \epsilon_{\alpha \beta \gamma \delta} \,
    p_{\mu}^{\alpha} \, p_{e}^{\beta} \,
    s_{\mu}^{\gamma} \, s_{e}^{\delta} \bigg).
 \end{Eqnarray}

 \noindent As usual $ x = 2 E_e / m_{\mu} $ denotes the scaled energy
 of the electron where the energy of the electron is defined in the rest
 frame of the $ \mu^{-} $. 
 $ \theta_P $ is the polar angle between the polarization of the muon
 and the momentum direction of the electron in the muon rest frame.

 Eq.~(\ref{invrate}) will be evaluated in the rest system of the muon
 where $ p_{\mu} = (m_{\mu}; 0, 0, 0) $ and 
 $ p_{e} = (E_{e}; 0, 0, |\vec{p_e}|) = $
 $ (m_{\mu}/2) (x; 0, 0, x \beta) $.
 The velocity of the electron is denoted by
 $ \beta = \sqrt{1 - 4 y^2 / x^2} $ where
 $ y = m_{e} / m_{\mu} $.
 In the rest frame of the $ \mu^{-} $ the polarization
 four--vectors of the $ \mu^{-} $ and $ e^{-} $ are given by

 \vspace{-5mm}
 \begin{Eqnarray} 
    \label{polarizationvectors}
    s_{\mu}^{\alpha} & = &
      (0; \vec{\zeta}_{\mu}), \\[3mm]
    s_{e}^{\alpha} & = &
      (\frac{\vec{n}_{e} \!\cdot\! \vec{p}_e}{m_{e}};
       \vec{n}_{e} +
       \frac{\vec{n}_{e} \!\cdot\! \vec{p}_{e}}{m_{e} (E_{e} + m_{e})}
       \vec{p}_{e}),
 \end{Eqnarray}
 \vspace{-3mm}
 
 \noindent where the polarization three--vector
 $ \vec{\zeta}_{\mu} $ of the $ \mu^{-} $
 and the quantization axis $ \vec{n}_{e} $ of
 the spin of the $ e^{-} $ in their respective
 rest frames read (see Fig.~1)

 \begin{equation} 
     \label{restframepolmu}
     \vec{\zeta}_{\mu} = (\sin \theta_P, \, 0, \, \cos \theta_{P})  
 \end{equation}
 
 \noindent and
 
 \begin{equation} 
     \label{restframepolelectron}
     \vec{n}_{e} =
     (\sin \theta \cos \chi, \, \sin \theta \sin \chi, \, \cos \theta).  
 \end{equation}

 Eq.~(\ref{restframepolmu}) holds for $ 100 \% $ polarized muons.
 For partially polarized muons with magnitude of polarization $ P $ the
 representation (\ref{restframepolmu}) has to be multiplied by $ P $ such
 that $ \vec{P}_{\mu} = P \vec{\zeta}_{\mu} $.
 The representation Eq.~(\ref{restframepolelectron})
 needs a word of explanation.
 The vector $\vec{n}_{e} $ denotes the orientation of the quantization
 axis of the electron's spin in the rest frame of the electron which can
 be freely chosen.
 The orientation of $ \vec{n}_{e} $ has no physical meaning per se.
 In particular $ \vec{n}_{e} $ is {\sl not} the polarization vector
 $ \vec{P}_{e} $ of the electron whose Cartesian components can, however,
 be obtained by projecting onto the
 $ x $--axis $ (\theta = \pi/2, \chi = 0) $, the 
 $ y $--axis $ (\theta = \pi/2, \chi = \pi/2) $
 and the $ z $--axis $ (\theta = 0) $.

 One finally has \cite{fs74,ms79}
 
 \begin{Eqnarray}   
   \label{diffrate} 
   \frac{d \Gamma}{dx \, d \! \cos \theta_P } & = &
   \beta x \, \Gamma_0 (G_1 + G_2 P \cos \theta_P +
    G_3 \cos \theta + G_4 P \cos \theta_P \cos \theta
   \nonumber \\[3mm] & + &
    G_5 P \sin \theta_P \sin \theta \cos \chi +
    G_6 P \sin \theta_P \sin \theta \sin \chi ).
 \end{Eqnarray}
 
 The relation between the invariant structure functions
 $ A_i $ and the frame--dependent spectrum functions
 $ G_i $ is given by

 \begin{eqnarray}     
   \label{conversion} 
    G_1 & = & \pp A_1, \nonumber \\[2mm]
    G_2 & = &  -  \frac{1}{2} x \beta A_2, \nonumber \\[2mm]
    G_3 & = & \pp \frac{1}{2 y} x \beta A_3, \nonumber \\[2mm]
    G_4 & = &  -  \frac{1}{4 y} x^2 \beta^2 A_4 -
                  \frac{1}{2 y} x A_5, \nonumber \\[2mm]
    G_5 & = &  -  A_5, \nonumber \\[2mm]
    G_6 & = & \pp \frac{1}{2} x \beta A_6.
 \end{eqnarray}

 \noindent $ G_1 $ is the unpolarized spectrum function,
 $ G_2 $ and $ G_3 $ are single spin polarized spectrum functions
 referring to the spins of the $ \mu^{-} $ and $ e^{-} $, resp.,
 and $ G_4 $, $ G_5 $ and $ G_6 $ describe
 spin--spin correlations between the spin vectors
 of the muon and electron\footnote{
 The spectrum function $ G_5 $ describing the transverse
 polarization of the electron vanishes for vanishing
 electron mass and is therefore not included in the
 corresponding decay distribution in \cite{fs74}.}.
 $ G_6 $ represents a so--called $ T $--odd observable.
 This is evident when rewriting the angular factor
 multiplying $ G_6 $ in (\ref{diffrate}) in triple--product
 form, i.e. $ \sin \theta_P \sin \theta \sin \chi = $
 $ |\vec{p}_e|^{-1} \vec{p}_e \cdot (\vec{\zeta}_\mu \times \vec{n}_e) $
 \footnote{A general discussion of the electron polarization in muon decay 
 can be found in \cite{ks57}, including a discussion of
 tests of the $TCP$ theorem.}.
 $ G_6 $ is identically zero in the SM since, on the one hand,
 the weak coupling constant $ G_F $ is real and, on the other hand,
 the loop contributions do not generate imaginary parts.
 $ G_6 $ will therefore not be discussed any further in the following.

 It is quite instructive to rewrite
 Eq.~(\ref{diffrate}) in a factorized form

 \begin{equation} 
    \frac{d \Gamma}{dx \, d \! \cos \theta_P } = 
    \beta x \, \Gamma_0 (G_1 + G_2 P \cos \theta_P) \,
    (1 + \vec{P}_{e} \cdot \vec{n}_{e}) 
 \end{equation}

 \noindent A comparison with Eq.~(\ref{diffrate}) shows that
 the polarization vector $ \vec{P}_{e} $ of the electron has the components

 \begin{Eqnarray}  
    \label{polvec} 
    P_{e}^{x} := P_{e}^{\perp} & = &
    \frac{P \, G_5 \sin \theta_P}
    {G_1 + G_2 P \cos \theta_P},
    \nonumber \\[3mm] 
    P_{e}^{y} := P_{e}^{N} & = &
    \frac{ P \, G_6 \sin \theta_P}
    {G_1 + G_2 P \cos \theta_P},
    \nonumber \\[3mm]
    P_{e}^{z} := P_{e}^{l} \; & = &
    \frac{G_3 + G_4 P \cos \theta_P}
    {G_1 + G_2 P \cos \theta_P}.
 \end{Eqnarray}

 \noindent We have as usual denoted the $ (x, y, z) $ components
 of $ \vec{P}_{e} $ by $ (P_{e}^{\perp}, P_{e}^{N}, P_{e}^{l}) $
 where $ (\perp, N, l ) $ stand for the polarization components
 transverse to the momentum direction of the electron (in the plane
 spanned by the momentum of the electron and the polarization vector
 of the muon), normal to this plane and longitudinal, respectively.
 Eq.~(\ref{polvec}) shows that the spectrum functions $ G_3 $ and
 $ G_4 $ determine the longitudinal polarization of the electron while
 its tranverse polarization (in the plain spanned by the electron's
 momentum and the muon's polarization) is determined by $ G_5 $\footnote{
 As noted before in the SM there is no transverse polarization normal
 to that plane.}.

 The limits on $ x  = 2 E_{e} / m_{\mu} $ are given by
 $ x_{min} = 2 y $ and $ x_{max} = 1 + y^2 $ where, as before,
 $ y = m_{e} / m_{\mu} $.
 $ \Gamma_0 $, finally, is the $ m_e = 0 $ Born term rate
 given by $ \Gamma_0 = G_F^2 m_{\mu}^5 / 192 \pi^3 $.
 The differential rate for the charge conjugated decay
 $ \mu^{+} \rightarrow e^{+} + \nu_{e} + \bar{\nu}_{\mu} $
 is obtained from Eq.~(\ref{diffrate}) by the substitution
 $ G_i \rightarrow G_i \: (i=1,4,5,6) $ and 
 $ G_i \rightarrow - G_i \: (i=2,3) $ \cite{ks57}.

 It is convenient to split the unpolarized and polarized rate functions
 into a Born term  part and a radiatively corrected  part according to

 \begin{equation} 
   \label{split}  
    G_i = G_i^{Born} + G_i^{(\alpha)} \hspace{1cm} i = 1,2,3,4,5 \, .
 \end{equation}

 \noindent The respective results on the Born term contributions
 $ G_i^{Born} $ and the $ O(\alpha) $ corrections to the rate functions
 $ G_i^{(\alpha)} $ are given in Secs.~3 and 5.
 The sum of the two contributions in Eq.~(\ref{split}) will generally
 be referred to as the next--to--leading order (NLO) result.

 
\section{Born term results}

 We shall work with the charge retention form of the Lagrangian for
 the decay $ \mu^{-} \rightarrow e^{-} + \bar{\nu}_{e} + \nu_{\mu} $
 which reads\footnote{The charge retention form of the Lagrangian is
 obtained from the usual Standard Model charged current--current form
 through the Fierz transformation (\ref{fierz1}) written down in
 Appendix B.
 The minus sign from the Fierz identity is cancelled from having to
 commute the Fermion fields an odd number of times in order to relate
 the two forms.
 Note that the Fierz identity is a four--dimensional identity.
 Since our calculations are done in four dimensions the Fierz identity
 can be safely applied.}
 $ { }^{,} $\footnote{When the Lagrangian for $ \mu $--decay is written
 in charge retention form the similarity of the decay
 $ \mu^{-} \rightarrow e^{-} + \bar{\nu}_{e} + \nu_{\mu} $ to the decay
 $ b \rightarrow c + e^{-} + \bar{\nu}_{e} $ becomes quite apparent
 through the substitutions $ b \leftrightarrow  \mu^{-} $,
 $ c \leftrightarrow e^{-} $, $ e^{-} \leftrightarrow \nu_{\mu} $ and
 $ \bar{\nu}_{e} \leftrightarrow \bar{\nu}_{e} $.}

 \begin{equation} 
   \label{chargeretention}
   {\cal L}(x) = \frac{G_F}{\sqrt{2}}
      \bar{\Psi}_{e}(x) \gamma^{\alpha}
      (\1 - \gamma_5) \Psi_{\mu}(x)
      \bar{\Psi}_{\nu_{\mu}}(x) \gamma_{\alpha}
      (\1 - \gamma_5) \Psi_{\nu_{e}}(x) + h.c. 
 \end{equation}

 \noindent When squaring the corresponding matrix element one obtains
 the tensor $ C^{\alpha \beta} $ from the charged lepton side
 ($ C $ for charged) which has to be contracted  with the neutrino--side
 tensor $ N^{\alpha \beta} $ ($ N $ for neutral).
 For the Born term contribution of the charge--side tensor one obtains

 \begin{equation} 
   \label{born}
    C_{Born}^{\alpha \beta} =
      \frac{1}{4} \mbox{Tr} \Big\{
      (\slp_{e} + m_{e})
      (\1 + \gamma_5 \sls_{e})
      \gamma^{\alpha}
      (\1 - \gamma_5)
      (\slp_{\mu} + m_{\mu})
      (\1 + \gamma_5 \sls_{\mu})
      \gamma^{\beta}
      (\1 - \gamma_5) \Big\},
 \end{equation}

 \noindent where the dependence on the polarization four--vectors of the
 $ \mu^{-} $ and $ e^{-} $ has been retained.

 Since only even--numbered $ \gamma $--matrix strings survive between the
 two $ (\1 - \gamma_5) $--factors in Eq.~(\ref{born}) one can compactly
 write the result of the trace evaluation as

 \begin{equation} 
   \label{bornspur}
    C_{Born}^{\alpha \beta} = 2 (
      \bar{p}_{\mu}^{\beta} \, \bar{p}_{e}^{\alpha} +
      \bar{p}_{\mu}^{\alpha} \, \bar{p}_{e}^{\beta} -
      g^{\alpha \beta} \, \bar{p}_{\mu} \!\cdot\! \bar{p}_{e} +
      i \epsilon^{\alpha \beta \gamma \delta}
      \bar{p}_{e,\gamma} \bar{p}_{\mu,\delta}),
 \end{equation}

 \noindent where

 \vspace{-6mm}
 \begin{Eqnarray} 
   \bar{p}_{\mu}^{\alpha} & = &
      p_{\mu}^{\alpha} -
      m_{\mu} s_{\mu}^{\alpha}, \\[4mm]
   \bar{p}_{e}^{\alpha} & = &
      p_{e}^{\alpha} -
      m_{e} s_{e}^{\alpha}
 \end{Eqnarray}
 \vspace{-3mm}
 
 \noindent and where $ s_{\mu}^{\alpha} $ and $ s_{e}^{\alpha} $ are the
 polarization four--vectors of the $ \mu^{-} $ and $ e^{-} $, resp.,
 defined in Eq.~(\ref{polarizationvectors}).

 The dependence on the momentum directions of the $ \bar{\nu}_{e} $--
 and $ \nu_{\mu} $--neutrinos has been completely integrated out in the
 differential rate.
 Thus the neutrino--side of the interaction can only depend on the spatial
 piece of the second rank tensor build from the momentum transfer to the
 neutrinos (for the present purpose the neutrinos are treated as massless) 
 which we denote by $ Q^{\alpha} $ \footnote{We denote the momentum transfer
 to the neutrino pair by a capital $ Q $ in order to set it apart from the
 momentum transfer $ q $ to the ($ e^{-} \bar{\nu}_{e} $)--pair used in
 Sec.~4.}. 
 Thus the relevant neutral--side tensor is given by

 \begin{equation} 
   \label{neutrinotensor}
    N^{\alpha \beta} = - g^{\alpha \beta} + \frac{Q^{\alpha} Q^{\beta}}{Q^2}.
 \end{equation}

 \noindent The Born spectrum functions can then be extracted from

 \vspace{0.5mm}
 \begin{equation} 
   \label{bornscalar}
    Q^2 N^{\alpha \beta} C^{Born}_{\alpha \beta} =
      2 ((\bar{p}_{\mu} \!\cdot\! \bar{p}_{e}) \, Q^2 + 
      2 (\bar{p}_{\mu} \!\cdot\! Q) \, (\bar{p}_{e} \!\cdot\! Q)),
 \end{equation}
 \vspace{0.5mm}
 
 \noindent where the antisymmetric piece in $ C^{Born}_{\alpha \beta} $
 has dropped out after the symmetric contraction.
 At the Born term level one has $ Q = p_{\mu} - p_{e} $.

 Including the correct normalization the differential Born term rate is
 given by

 \begin{equation} 
   \label{diffbornrate}
   \frac{d \Gamma^{Born}}{dx \, d \! \cos \theta_P } = \Gamma_0 \, \beta x
   \frac{Q^2 N^{\alpha \beta} C^{Born}_{\alpha \beta}}{m_{\mu}^4}.
 \end{equation}

 \noindent The spectrum functions defined in Eq.~(\ref{diffrate})
 can then easily be calculated from the Born term contributions
 (\ref{bornscalar}) using the relations Eq.~(\ref{conversion}).
 They are given by

 \begin{equation} 
   \label{bornratefunctions}
   \begin{array}{rcl}
      G_1^{Born} & = & \pp x (3 - 2 x) - (4 - 3 x) y^2, \\[4mm]
      G_2^{Born} & = & \pp \beta x (1 - 2 x + 3 y^2), \\[4mm]
      G_3^{Born} & = &  -  \beta x (3 - 2 x + y^2), \\[4mm]
      G_4^{Born} & = &  -  x (1 - 2 x) - (4 + x) y^2, \\[4mm]
      G_5^{Born} & = &  -  2 y (1 - x + y^2).
 \end{array}
 \end{equation}
 
 \noindent Note that the Born term spectrum functions
 $ G_{1, 2, 3, 4}^{Born} $ and $ G_5^{Born} / y $ are
 quadratically dependent on $ y $.
 This is in agreement with the general arguments
 presented in \cite{roos71}.
 For $ y^2 = 0 \, (m_e = 0) $ one has $ G_1^{Born} = - G_3^{Born} $
 and $ G_2^{Born} = - G_4^{Born} $ which reflects the fact that a
 mass zero left--chiral electron is purely lefthanded.
 The Born term results (\ref{bornratefunctions}) reproduce the
 $ y = 0 $ results of Ref.~\cite{fs74}.
 For $ y^2 \ne 0 \, (m_e \ne 0) $ our Born term results for
 $ G_1^{Born} $ and $ G_2^{Born} $ agree with those of
 Ref.~\cite{arbuzov01}.
 Note that the spectrum function $ G_5^{Born} $ is proportional
 to $ y $ and thus vanishes for vanishing electron mass.
 The overall chiral factor $ y = m_{e} / m_{\mu} $ in $ G_5 $
 originates from a lefthanded/righthanded interference contribution
 which is chirally suppressed.

 In Figs.~2a ($ \mu \rightarrow e $), 3a ($ \tau \rightarrow \mu $)
 and 4a ($ \tau \rightarrow e $) we show plots of the $ x $--dependence
 of the four Born term spectrum functions $ \beta x G_{1, 2, 3, 4}^{Born} $.
 They rise and fall from zero at the soft end of the spectrum to
 $ (1 - y^2)^3 \, (i = 1,4) $ and $ - (1 - y^2)^3 \, (i = 2,3) $
 at the hard end of the spectrum, respectively.
 The $ m_{l'} = 0 $ pattern $ G_1^{Born} = - G_3^{Born}$ and
 $ G_2^{Born} = - G_4^{Born} $ is slightly distorted by final
 lepton mass effects except for the point $ x_{max} = 1 + y^2 $
 where the above $ m_{l'} = 0 $ relations are exact.
 For the spectrum function $ G_5^{Born} $ one finds
 $ G_5^{Born}(x_{min} = 2 y) = - 2 y (1 - y)^2 $ and
 $ G_5^{Born}(x_{max} = 1 + y^2) = 0 $.
 The chirally suppressed contribution of $ \beta x G_5^{Born} $
 is negative over the whole $ x $--range.
 At the scale of the plots it is only visible for the case
 $ \tau \rightarrow \mu $ (Fig.~3a).

 In addition to the polarization observables we also define a
 forward--backward asymmetry for the case that the spin of the
 electron is not observed.
 It reads

 \begin{equation}  
   \label{forwardbackward1}
   A_{FB} = \frac{\Gamma_F - \Gamma_B}{\Gamma_F + \Gamma_B}
          = \frac{1}{2} \; P \; \frac{G_2}{G_1}, 
 \end{equation}

 \noindent where $ \Gamma_F $ and $ \Gamma_B $ are the rates into the forward
 ($ \cos \theta_P \ge 0 $) and backward ($ \cos \theta_P \le 0 $) hemispheres.

 One can also define a forward--backward asymmetry of the longitudinal
 polarization proportional to $ G_4 / G_1 $ according to

 \begin{equation}              
   \label{forwardbackward1ong} 
   P^l_{l'\, (F\!B)} = \frac{
     \Gamma_F^{+} - \Gamma_F^{-} - \Gamma_B^{+} + \Gamma_B^{-}}{
     \Gamma_F^{+} + \Gamma_F^{-} + \Gamma_B^{+} + \Gamma_B^{-}} =
     \frac{1}{2} \; P \; \frac{G_4}{G_1}, 
 \end{equation}

 \noindent where the indices $ +/- $ denote the helicities of the electron.
 This asymmetry will be difficult to measure since it involves a spin--spin
 correlation measurement.
 We shall therefore not discuss this asymmetry any further in this paper.
 
 The longitudinal polarization and the forward--backward asymmetry
 take the values $ P_{e}^{l, Born} = - \frac{1}{3} P \cos \theta_P $
 and $ A_{F\!B}^{Born} = 0 $, respectively, at the lower limit
 $ x_{min} = 2 y $ where $ G_1^{Born} = - 3 G_4^{Born} $ and
 $ G_2^{Born} = G_3^{Born} = 0 $.
 This has to be contrasted with the naive limits $ P_{e}^{l \, Born} = - 1 $
 and $ A_{F\!B}^{Born} = \frac{1}{6} P $ when naively setting $ y = 0 $ in
 the corresponding ratios.
 At the upper limit $ x_{max} = (1 + y^2) $, where 
 $ G_1^{Born} = - G_2^{Born} = - G_3^{Born} = G_4^{Born} = (1 - y^2)^2 $,
 the longitudinal polarization decreases to
 $ P_e^{l \, Born } = - 1 $ irrespective of
 the value of $ P \cos \theta_P $, and the
 forward--backward asymmetry decreases to 
 $ A_{FB}^{Born}= - \frac{1}{2} P $ as in
 the naive $ m_e = 0 $ case.
 At the soft end of the spectrum one finds a substantial transverse
 polarization $ P_{e}^{\perp, Born} = - \frac{1}{3} P \sin \theta_P $
 which is independent of $ y $ contrary to naive expectations.
 Thus the total polarization of the electron at threshold is given by
 $ |\vec{P_e}^{Born}| = \frac{1}{3} P $ irrespective of the value of
 $ \cos \theta_P $. 
 At the hard end of the spectrum one has $ P_{e}^{\perp, Born} = 0 $.

 In Figs.~2b ($ \mu \rightarrow e $), 3b ($ \tau \rightarrow \mu $) and
 4b ($ \tau \rightarrow e $) we show the $ x $--dependence of the Born term
 prediction for the longitudinal and transverse polarization of the
 daughter lepton.
 We set $ P = 1 $ and take three values $ \cos \theta_P = 1, 0 $ and $ -1 $
 for the longitudinal polarization and set $ \cos \theta_P = 0 $ for the
 transverse polarization.
 The longitudinal polarization stays close to $ -1 $ over most of the
 (hard part) of the spectrum but deviates significantly from the naive
 value $ -1 $ in the threshold region with only a slight dependence on
 the value of $ \cos \theta_P $.
 In order to highlight the deviations from the naive value
 $ P_{l'}^{l, Born} = -1 $ in the threshold region we have chosen
 logarithmic scales for the energy variable $ x $.
 The transverse polarization is negative and stays very close to
 zero over most of the (hard part) of the spectrum and decreases to
 its limiting value $ - 1/3 $ at threshold.
 In fact all the Born term curves can be seen to approach the limits
 discussed above at the soft and hard end of the spectrum.

 In Figs.~2c, 3c and 4c we show the corresponding curves for the
 forward--backward asymmetry $ A_{F\!B}^{Born} $ for the three decay cases.
 Again we set $ P = 1 $.
 The forward--backward asymmetries rise from the limiting value
 $ A_{F\!B}^{Born} = 0 $ at threshold  and then fall to
 $ A_{F\!B}^{Born} = - 1/2 $ at the hard end of the spectrum.

 \renewcommand{\arraystretch}{1.3}
 \begin{table}
 \rule{160mm}{1pt}
 \begin{tabular*}{160mm}{l@{\extracolsep{\fill}}|ll|ll|ll}
  \hline
  & \multicolumn{2}{c|}{$ \mu  \rightarrow  e  $} 
  & \multicolumn{2}{c|}{$ \tau \rightarrow \mu $}
  & \multicolumn{2}{c }{$ \tau \rightarrow  e  $} \\
  & Born & $ \quad  (\alpha) $
  & Born & $ \qquad (\alpha) $
  & Born & $ \qquad (\alpha) $ \\
  \hline
   $ \hat{G}_{1} \, $ & $ - 0.019 \% $ & $ - 0.12 \% $ &
   $ - 2.82 \% $ & $ - \phantom{1} 8.48 \% $ &
   $ - 6.62 \cdot 10^{-5} \% $ & $ - 7.39 \cdot 10^{-4} \% $ \\
   $ \hat{G}_{2} \, $ & $ - 3.57 \cdot 10^{-4} \, \% $ & $ - 0.67 \% $ &
   $ - 0.57 \% $ & $ - 11.23 \% $ & 
   $ - 7.60 \cdot 10^{-8} \% $ & $ - 0.039 \% $ \\
   $ \hat{G}_{3} \, $ & $ - 0.031 \% $ & $ - 1.48 \% $ &
   $ - 4.25 \% $ & $ - 24.49 \% $ &
   $ - 1.10 \cdot 10^{-4} \% $ & $ - 0.084 \% $ \\
   $ \hat{G}_{4} \, $ & $ - 0.019 \% $ & $ - 1.43 \% $ &
   $ - 2.82 \% $ & $ - 22.15 \% $ &
   $ - 6.62 \cdot 10^{-5} \, \% $ & $ - 0.083 \% $ \\
  \hline
 \end{tabular*}
 \rule{160mm}{1pt} \\[0.8ex]
 \centerline{\bf Table 1:}
 {\small Percentage mass corrections to $ m_{l'} = 0 $ (Born) and
  $ m_{l'} \rightarrow 0 $ ($ O(\alpha) $) rate functions
  for the three cases ($ \mu \rightarrow e $),
  ($ \tau \rightarrow \mu $) and ($ \tau \rightarrow e $).}
 \end{table}

 Next we integrate the differential Born term rates
 over the full $ x $ spectrum.
 Let us define reduced Born rate functions
 $ \hat{G}_i^{Born} $ according to

 \begin{equation} 
   \label{reducedrate}
   \hat{G}_i^{Born} =
   \int\limits_{2 y}^{1 + y^2} \!\!\! dx \,
   \beta x \, G_i^{Born} \hspace{1cm} i = 1,2,3,4.
 \end{equation}

 \noindent One obtains

 \vspace{1mm}
 \begin{equation} 
   \label{intbornratefunctions}
   \begin{array}{r@{\,=\,}l}
     \hat{G}_{1}^{Born} & \pp
     \frac{1}{2}(1 - y^4)(1 - 8 y^2 + y^4) - 12 y^4 \ln y =
     \frac{1}{2} - 4 y^2 + O(y^4), \\[4mm]
     \hat{G}_{2}^{Born} & -
     \frac{1}{6}(1 - y)^5 (1 + 5 y + 15 y^2 + 3 y^3) = -
     \frac{1}{6} + \frac{16}{3} y^3 + O(y^4),\\[4mm]
     \hat{G}_{3}^{Born} & -
     \frac{1}{6}(1 - y)^5 (3 + 15 y + 5 y^2 + y^3) = -
     \frac{1}{2} + \frac{20}{3} y^2 - 16 y^3 + O(y^4),\\[4mm]
     \hat{G}_{4}^{Born} & \pp
     \frac{1}{6}(1 - y^4)(1 - 8 y^2 + y^4) - 4 y^4 \ln y =
     \frac{1}{6} - \frac{4}{3} y^2 + O(y^4),\\[4mm]
     \hat{G}_{5}^{Born} & - y
     \big(\frac{1}{3} (1 - y^2)(1 + 10 y^2 + y^4) +
     4 y^2 (1 + y^2) \ln y \big) = - y
     \big( \frac{1}{3}  + (3 \!+\! 4 \ln y) y^2 + O(y^4)\big)
     \hspace*{-1.5cm}
  \end{array}
 \end{equation}
 \vspace{1mm}

 The occurrence of odd powers of $ y $ in $ \hat{G}_{2}^{Born} $ and
 $ \hat{G}_{3}^{Born} $ can be traced to the lower boundary $ x = 2 y $
 of the $ x $--integration which is linear in $ y $.
 In view of this it is quite remarkable that $ \hat{G}_{1}^{Born} $,
 $ \hat{G}_{4}^{Born} $ and $ \hat{G}_{5}^{Born} / y $ contain only
 even powers of $ y $.
 Nevertheless, the leading mass correction to
 $ \hat{G}_{3}^{Born} $ sets in only at $ O(y^2) $ and, for
 $ \hat{G}_{2}^{Born} $, only at  $ O(y^3) $.
 In this sense the Born term rates are well protected against
 finite electron mass effects, or, in the case
 $ (\tau \rightarrow \mu) $, reasonably well against muon mass effects.
 This is illustrated in Table~1 where we list the values of the Born term
 percentage changes $ (\hat{G}_{i} (m \ne 0) - \hat{G}_{i} (m = 0)) $
 $ / \hat{G}_{i} (m = 0) $ when going from $ m_{l'} = 0 $ to
 $ m_{l'} \ne 0 $ for all three cases.
 
 The average longitudinal polarization of the electron
 $ \langle P_{e}^{l} \rangle $ and the average forward--backward
 asymmetry $ \langle A_{F\!B} \rangle $ is obtained by the
 replacement ($ G_i \rightarrow  \hat{G}_i $) in
 (\ref{polvec}) and (\ref{forwardbackward1}).
 As in the rate expressions (\ref{intbornratefunctions}) the
 final state lepton mass effects on the Born term polarization
 $ \langle P_{e}^{l} \rangle $ and forward--backward
 asymmetry $ \langle A_{F\!B} \rangle $ are quite small.
 The deviation from the $ m_{l'} = 0 $ result
 $ \langle P_{l'}^{l} \rangle = -1 $ is of
 $ O(10^{-4}) $ in muon decay and $ O(10^{-6}) $ in
 $ (\tau \rightarrow e) $.
 For $ (\tau \rightarrow \mu) $ the deviation from
 $ \langle P_{\mu}^{l} \rangle = -1 $ is of $ O(10^{-2}) $.
 The dependence on the value of $ P \cos\theta_P $ is very small.
 The average value of the forward--backward asymmetry
 $ \langle A_{F\!B} \rangle $ is quite close
 ($ O(10^{-2} - 10^{-3}) $) to the $ y = 0 $ prediction
 $ \langle A_{F\!B} \rangle = - 1/6 $.
 The average transverse polarization
 $ \langle P_{\mu}^{\perp} \rangle $ is generally quite small
 due to the overall chiral factor $ y = m_{l'} / m_{l} $.
 To be specific, for $ \theta_P = \pi/2 $ one
 finds an average transverse polarization of $ - 0.3 \% $,
 $ - 3.7 \% $ and $ - 0.02 \% $ for the three cases
 $ (\mu \rightarrow e) $, $ (\tau \rightarrow \mu) $
 and $ (\tau \rightarrow e) $, respectively.


\section{\protect \boldmath $ W $-boson propagator effects }

 In order to incorporate $ W $--boson propagator effects one has to
 rewrite the charge retention form of the four--Fermi interaction
 (\ref{chargeretention}) in terms of the charged currents of the
 Standard Model including the $ W $--boson propagator.
 This is easily done using the Fierz transformation property
 written down in Eq.~(\ref{fierz1}) in Appendix~B. 

 For the present purposes it is only necessary to take into account
 $ W $--boson propagator effects in the Born term contributions.
 The momentum transfer is now
 $ q = p_{\mu} - p_{\nu_{\mu}} = p_e + p_{\nu_e} $.
 The $ q^{\mu} q^{\nu} $ piece of the $ W $--boson propagator
 contributes only at $ O(m_e^2 m_{\mu}^2 / m_W^4) $
 in the spectrum and rate functions and can therefore be dropped.
 In fact, using the Fierz identity (\ref{fierz2}), it is not difficult
 to compute the contribution of the $ q^{\mu} q^{\nu} $ piece exactly.
 The $ W $--boson propagator effect on the spectrum can thus be taken
 into account by the replacement
 
 \begin{eqnarray} 
   \label{spectrumpropagator}
    1 \, \rightarrow \, \left(
    \frac{m_W^2}{q^2 - m_W^2} \right)^2 \,
    & \approx & \, 1 + \frac{m_{\mu}^2}{m_W^2}
    \frac{x (2 - x)}{3 - 2 x} \quad i = 1, 3 \quad, \\
    & \approx & \, 1 - \frac{m_{\mu}^2}{m_W^2}
    \frac{x^2}{1 - 2 x} \qquad  i = 2, 4 \quad. 
 \end{eqnarray}

 \noindent where terms of $ O(m_e^2 / m_W^2) $ have been neglected.
 Numerically the propagator corrections are quite small since
 $ m_{\mu}^2  / m_W^2 = 1.73 \cdot 10^{-6} $ and
 $ m_{\tau}^2 / m_W^2 = 4.88 \cdot 10^{-4} $.

 As is evident from Eq.~(\ref{spectrumpropagator}) the $ W $--boson
 propagator affects the two pairs of spectrum functions differently.
 This means that one cannot absorb the $ W $--boson propagator effect
 entirely into a redefinition of Fermi's coupling constant $ G_F $,
 as advocated in Refs.~\cite{hagiwara02,marciano88}, when one considers
 polarization effects.

 In order to determine the propagator corrections to the
 rate functions one has to do the integrations

 \begin{equation} 
   \label{ratepropagator1}
   \int_{0}^{1} \!\! dx \, x \, G_{1,3}^{Born}
   (1 +  \frac{m_{\mu}^2}{m_W^2} \frac{x (2 - x}{3 - 2 x}) =
   \pm \frac{1}{2} (1 + \frac{3}{5} \frac{m_{\mu}^2}{m_W^2}),
 \end{equation}

 \begin{equation} 
  \label{ratepropagator2}
   \int_{0}^{1} \!\! dx \, x \, G_{2,4}^{Born}
   (1 - \frac{m_{\mu}^2}{m_W^2} \frac{x^2}{1 - 2 x}) =
   \mp \frac{1}{6} (1 + \frac{1}{5} \frac{m_{\mu}^2}{m_W^2}),
 \end{equation}

 \noindent where terms of $ O(y) $ have been dropped in the
 Born term factors and in the integration measure.
 Again it is evident from Eq.~(\ref{ratepropagator1}) and
 Eq.~(\ref{ratepropagator2}) that the $ W $--boson propagator
 affects the two pairs of rate functions differently.
 If the $ W $--boson propagator effect is absorbed into a
 redefinition of $ G_F $ using a measurement of the total
 unpolarized rate then this must be compensated for by
 multiplying the polarization dependent pieces $ \hat{G}_{2,4} $
 by $ (1 - \frac{2}{5} \frac{m_{\mu}^2}{m_W^2}) $ when
 calculating the average of the longitudinal polarization of the
 electron $ \langle P_e^l \rangle $ or the forward--backward
 asymmetry $ \langle A_{F\!B} \rangle $.
 
 
\section{\protect \boldmath $ O(\alpha) $ corrections to
 spin dependent rate functions}

 Many of the technical ingredients that go into the calculation of the
 $ O(\alpha) $ corrections can be found in a detailed account of the
 $ O(\alpha_s) $ corrections to the decay of a polarized top into a bottom
 quark and a $ W $ gauge boson $ t \rightarrow b + W^{+} $ presented in
 Ref.~\cite{fgkm01} (for the loop contribution see also \cite{schilcher}).
 Compared to Ref.~\cite{fgkm01} one needs to include the polarization
 dependence of the final massive fermion in this application.
 For the $ O(\alpha) $ tree--graph and one--loop contributions this is
 easily done.

 Let us briefly discuss the tree graph contribution.
 The $ O(e) $ tree graph amplitude (``internal bremsstrahlung'')
 consists of the two contributions where the photon is either
 radiated off the electron or off the muon.
 One thus has

 \begin{equation} 
    {\cal M}^{\alpha} = e \,\bar{u}_e \left(
    \gamma_{\delta}
    \frac{\slp_e + \slk + m_e}{(p_e + k)^2 - m_e^2}
    \gamma^{\alpha} (\1 - \gamma_5) +
    \gamma^{\alpha} (\1 - \gamma_5)
    \frac{\slp_{\mu} - \slk + m_{\mu}}{(p_{\mu} -k)^2 - m_{\mu}^2}
    \gamma_{\delta}
    \right) u_{\mu} \, \epsilon^{\ast \,\delta},
 \end{equation}

 \noindent where $ k $ and $ \epsilon_{\delta} $ are the momentum
 and the  polarization four--vector of the photon.
 Four--momentum conservation now reads $ p_{\mu} = p_e + Q + k $
 where $ Q $ is again the momentum transferred to the neutrino pair.
 When squaring the tree graph amplitude one only sums over
 the polarization states of the photon since the muon and
 the electron are taken as polarized.
 Omitting again the antisymmetric contribution
 one obtains in the Feynman gauge

 \begin{Eqnarray} 
  \label{matrix-element}
   C^{ ( \alpha )\alpha \beta} & = & \sum_{\gamma-spin}
   {\cal M}^{\alpha} {\cal M}^{\beta \dagger} =
  \frac{e^2}{2} \Bigg\{ \frac{1}{(k \!\cdot\! p_{e})} \Bigg(
  \frac{k \!\cdot\! \bar{p}_{e} - m_{e}^2}{(k \!\cdot\! p_{e})} +
  \frac{p_{\mu} \!\cdot\! \bar{p}_{e}}{(k \!\cdot\! p_{\mu})} \Bigg)
  (k^{\alpha} \bar{p}_{\mu}^{\beta} + k^{\beta} \bar{p}_{\mu}^{\alpha} -
   k \!\cdot\! \bar{p}_{\mu} g^{\alpha \beta}) \nonumber \\[2mm] & + &
  \frac{1}{(k \!\cdot\! p_{\mu})} \Bigg(
  \frac{k \!\cdot\! \bar{p}_{\mu} + m_{\mu}^2}{(k \!\cdot\! p_{\mu})} -
  \frac{p_{e} \!\cdot\! \bar{p}_{\mu}}{(k \!\cdot\! p_{e})} \Bigg)
  (k^{\alpha} \bar{p}_{e}^{\beta} + k^{\beta} \bar{p}_{e}^{\alpha} -
   k \!\cdot\! \bar{p}_{e} g^{\alpha \beta}) \nonumber \\[2mm] & + &
  \frac{k \!\cdot\! \bar{p}_{e}}{(k \!\cdot\! p_{e})^2}
  (p_{e}^{\alpha} \bar{p}_{\mu}^{\beta} +
   p_{e}^{\beta} \bar{p}_{\mu}^{\alpha} -
   p_{e} \!\cdot\! \bar{p}_{\mu} g^{\alpha \beta}) -
  \frac{k \!\cdot\! \bar{p}_{\mu}}{(k \!\cdot\! p_{\mu})^2}
  (p_{\mu}^{\alpha} \bar{p}_{e}^{\beta} +
   p_{\mu}^{\beta} \bar{p}_{e}^{\alpha} -
   p_{\mu} \!\cdot\! \bar{p}_{e} g^{\alpha \beta}) \nonumber \\[2mm] & + &
  \frac{k \!\cdot\! \bar{p}_{\mu}}
  {(k \!\cdot\! p_{e})(k \!\cdot\! p_{\mu})}
  (p_{e}^{\alpha} \bar{p}_{e}^{\beta} +
   p_{e}^{\beta} \bar{p}_{e}^{\alpha} -
   m_{e}^2 g^{\alpha \beta}) -
  \frac{k \!\cdot\! \bar{p}_{e}}{(k \!\cdot\! p_{e})(k \!\cdot\! p_{\mu})}
  (p_{\mu}^{\alpha} \bar{p}_{\mu}^{\beta} +
   p_{\mu}^{\beta} \bar{p}_{\mu}^{\alpha} -
   m_{\mu}^2 g^{\alpha \beta}) \Bigg\} \nonumber \\[2mm] & - &
  \frac{e^2}{2} \Bigg(
  \bar{p}_{e}^{\alpha} \bar{p}_{\mu}^{\beta} +
  \bar{p}_{e}^{\beta} \bar{p}_{\mu}^{\alpha} -
  \bar{p}_{e} \!\cdot\! \bar{p}_{\mu} g^{\alpha \beta} \Bigg)
  \Bigg(
  \frac{m_{\mu}^2}{(k \!\cdot\! p_{\mu})^2} +
  \frac{m_{e}^2}{(k \!\cdot\! p_{e})^2} - 2
  \frac{p_{e} \cdot\! p_{\mu}}
  {(k \!\cdot\! p_{e})(k \!\cdot\! p_{\mu})}
  \Bigg).
 \end{Eqnarray}

 \noindent In the last line of Eq.~(\ref{matrix-element}) we have
 isolated the infrared singular piece of the charge--side tensor which
 is given by the usual soft photon factor multiplying the Born term
 contribution.
 In the phase space integration over the photon momentum the infrared
 singular piece is regularized by introducing a (small) photon mass
 which distorts the phase space away from the singular point.
 The infrared singular piece is cancelled by the corresponding
 singular piece in the one--loop contributions again calculated in
 Feynman gauge. In the one--loop amplitude the photon mass is
 included in the pole denominators only\footnote{
 We specify our infrared regularization procedure since, historically,
 there has been a certain amount of controversy concerning the use of
 a photon mass regulator (see e.g. \cite{kinoshita01}).}.

 The remaining part of the charge--side tensor in
 Eq.~(\ref{matrix-element}) is infrared finite and can
 easily be integrated without a regulator photon mass.

 Just as in the Born term case treated in Sec.~3 the
 $ O(\alpha) $ charge--side tensor $ C^{(\alpha)}_{\alpha \beta} $
 is contracted with the neutral--side tensor $ N^{\alpha \beta } $
 Eq.~(\ref{neutrinotensor}) where now $ Q = p_{\mu} - p_e - k $.
 In the tree graph integration one first integrates over the photon's energy
 and then over the cosine of the angle between the photon and the electron
 taking appropriate care of the infrared singular piece. Any azimuthally
 dependent terms drop out after the azimuthal integration. One then finally
 adds in the one--loop contributions. The final result is  

 \begin{Eqnarray} 
  \label{spectrum1}
   G_{1}^{(\alpha)} & = & \frac{\alpha}{\pi} G_1^{Born} \, \Sigma +
  \frac{\alpha}{\pi} \frac{1}{12 \beta x} \Bigg\{
   3 \beta x \Big( (3 - 4 x) x - (8 - 9 x) y^2 \Big)
   \lambda_{1} \nonumber \\[2mm] & + &
   \Big( (5 + 12 x - 51 x^2 + 28 x^3) - 3 (19 - 40 x + 17 x^2) y^2 -
   3 (19 - 4 x) y^4 + 5 y^6 \Big) \lambda_{2} \nonumber \\[2mm] & - &
   4 \beta x \Big( (11 - 10 x + 5 x^2) - 2 (1 + 5 x) y^2 + 11 y^4 \Big)
  \Bigg\} ,\\[3mm]
  \label{spectrum2} 
   G_{2}^{(\alpha)} & = & \frac{\alpha}{\pi} G_2^{Born} \, \Sigma +
  \frac{\alpha}{\pi} \frac{1}{12 \beta^2 x^2} \Bigg\{
   3 \beta^3 x^3 \Big(1 - 4 x + 9 y^2 \Big) \lambda_1 - 
   8 \Big( 1 - x + y^2 \Big)^3 \lambda_3 \nonumber \\[2mm] & - &
  \Big( x (1 \!+\! 33 x^2 \!-\! 28 x^3) -
   3 x (33 \!-\! 12 x \!-\! 17 x^2) y^2 +
   3 (32 - 35 x - 4 x^2) y^4 - 5 x y^6 \Big)
  \lambda_2 \nonumber \\[2mm] & - &
   2 \beta x \Big( (3 - 10 x - 13 x^2 + 8 x^3) + (59 - 11 x^2) y^2 -
  (35 - 10 x) y^4 + 5 y^6 \Big) \Bigg\}, \\[3mm]
  \label{spectrum3}
   G_{3}^{(\alpha)} & = & \frac{\alpha}{\pi} G_3^{Born} \, \Sigma + 
  \frac{\alpha}{\pi} \frac{1}{12 \beta^2 x^2} \Bigg\{ -
   3 \beta^3 x^3 (3 - 4 x + 3 y^2) \lambda_1 -
   8 y^2 \Big( 1 - x + y^2 \Big)^3 \lambda_3 \nonumber \\[2mm] & - &
  \Big( x (5 + 12 x - 51 x^2 + 28 x^3) -
   (8 - 129 x + 60 x^2 + 25 x^3) y^2 \nonumber \\[2mm] & - &
   3 (40 - 49 x + 8 x^2) y^4 - (24 - 23 x) y^6 -
   8 y^8 \Big) \lambda_2 \nonumber \\[2mm] & + &
   2 \beta x \Big( (5 + 10 x - 11 x^2 + 8 x^3) - (35 + 13 x^2) y^2 +
   (59 - 10 x) y^4 + 3 y^6 \Big) \Bigg\}, \\[3mm]
  \label{spectrum4}
   G_{4}^{(\alpha)} & = & \frac{\alpha}{\pi} G_4^{Born} \, \Sigma +
  \frac{\alpha}{\pi} \frac{1}{12 \beta^3 x^3} \Bigg\{ -
   3 \beta^3 x^3 \Big( (1 - 4 x) x + (8 + 3 x) y^2 \Big)
   \lambda_1 \nonumber \\[2mm] & + &
   8 x (1 - y^2) (1 - x + y^2)^3 \lambda_3 + 4 \beta x \Big(
   x (1 - 5 x - 5 x^2 + 3 x^3) \nonumber \\[2mm] & - &
   (4 - 59 x + 14 x^2 + 5 x^3) y^2 - (104 - 59 x + 5 x^2) y^4 -
   (4 - x) y^6 \Big) \nonumber \\[2mm] & + &
  \Big( x^2 (1 + 33 x^2 - 28 x^3) +
   (4 + 8 x - 153 x^2 + 104 x^3 + 25 x^4) y^2 \nonumber \\[2mm] & - &
   3 (28 - 72 x + 59 x^2 - 8 x^3) y^4 - (84 - 24 x + 23 x^2) y^6 +
   4 (1 + 2 x) y^8 \Big) \lambda_2 \Bigg\} \\[3mm]
  \label{spectrum5}
   G_{5}^{(\alpha)} & = & \frac{\alpha}{\pi} G_5^{Born} \, \Sigma +
  \frac{\alpha}{\pi} \frac{y}{2} \Bigg\{ -
  (1 - 2 x + 3 y^2) \lambda_1 + \frac{2}{\beta x}
  (3 x - 2 x^2 - (4 - 3 x) y^2) \lambda_2 \Bigg\}.
 \end{Eqnarray}

 \noindent We have used the abbreviations

 \begin{equation} 
   \lambda_1 = \ln y^2, \qquad
   \lambda_2 = \ln \Bigg( \frac{1 + \beta}{1 - \beta} \Bigg), \qquad
   \lambda_3 = \ln \Bigg( \frac{2 - (1 + \beta) x}{2 - (1 - \beta) x} \Bigg)
 \end{equation}

 \noindent and

 \begin{equation} 
 \label{sigma}
 \begin{array}{cl}
   \displaystyle \Sigma & = \,
   \displaystyle \frac{1}{\beta} \Bigg\{
    2 \Li \Bigg( \frac{2 \beta x}{2 - (1 - \beta) x} \Bigg) -
    2 \Li \Bigg( \frac{2 \beta x}{(1 + \beta) x - 2 y^2} \Bigg) -
   \beta \ln (1 - x + y^2) \\[9mm] & + \,
   \displaystyle \Bigg( \ln \! \Bigg((1 + \beta) \frac{x}{2} \Bigg) -
   \frac{1 - y^2}{x} \Bigg) (\lambda_3 - \lambda_2) +
   \Bigg( \ln \! \Bigg(1 - (1 + \beta) \frac{x}{2} \Bigg) -
   \frac{1 - x}{x} \Bigg) \lambda_2 \Bigg\}.
 \end{array}
 \end{equation}

 \noindent The radiative corrections to the unpolarized spectrum function $ G_1 $
 including the full mass dependence have first been calculated in \cite{bfs56}.
 After correcting for an error in the calculation of \cite{bfs56}, a correct result
 was published in \cite{ks59}. Our results for $ G_1^{(\alpha)} $ agree with those
 given in \cite{ks59} and in \cite{arbuzov01}. Our results for $ G_2^{(\alpha)} $
 agree with those of \cite{arbuzov01}. The results on $ G_{3,4,5}^{(\alpha)} $
 are new. 

 Note that that the four spectrum functions $ G_{1, 2, 3, 4}^{(\alpha)} $
 are logarithmically mass divergent.
 We have checked that the expansion of the spectrum functions
 $ G_{1, 2, 3, 4}^{(\alpha)} $ and $ G_5^{(\alpha)}/y $ 
 in terms of powers of $ y $ contains only even powers of $ y $,
 in agreement with the general reasoning given in \cite{roos71}.
 The leading term in the $ y $--expansion of the spectrum
 functions $ G_{1, 2, 3, 4}^{(\alpha)} $ can be reconstructed
 from the $ m_e \rightarrow 0 $ results to be presented in Sec.~6.
 We do not write down any of the higher order coefficients in the
 $ y $--expansion, as was done in the Born term case, since such
 an expansion is not particularly illuminating.

 The explicit evaluation of the five spectrum functions is numerically quite
 stable except for the region very close to threshold, i.e. for $\beta$--values
 smaller than $ \beta \approx 0.05 $, or, when expressed in terms of the scaled
 electron energy $ x $, for $ x $--values below $ (9.7 \times 10^{-3};
 1.2 \times 10^{-1} ; 5.8 \times 10^{-4}) $. The origin of the instability
 are the inverse powers of $ \beta $ appearing in 
 Eqs.(\ref{spectrum1}-\ref{spectrum5}). If one wants
 to explore the region below $ \beta \approx 0.05 $ one can make use of a small
 $ \beta $--expansion of the expressions Eqs.(\ref{spectrum1}-\ref{sigma})
 which is not difficult to arrive at using an algebraic program such as e.g.
 MATHEMATICA. We do not write down explicit forms for the small $ \beta $--expansion
 because the expressions are not particularly illuminating. Let it be said that
 in the small $ \beta $--expansion $ G_{1,4,5}^{(\alpha)} $ and
 $ G_{2,3}^{(\alpha)} $ are even and odd functions of $ \beta $, respectively. This
 shows that $ G_2^{(\alpha)} $ and $ G_3^{(\alpha)} $ are
 proportional to $ \beta $ just as in the Born term case. We mention that
 approximate formulas for the threshold region have been written down for $ G_1 $
 in \cite{grotch}.
 
 In Fig.~2a ($ \mu \rightarrow e $), Fig.~3a ($ \tau \rightarrow \mu $)
 and Fig.~4a ($ \tau \rightarrow e $) we show plots of the $ x $--dependence
 of the four spectrum functions $ \beta x G_i \,(i=1, 2, 3, 4) $, with and
 without radiative corrections.
 The radiative corrections show a marked $ y $--dependence.
 They are smallest for $ \tau \rightarrow \mu $, become larger for
 $ \mu \rightarrow e $, and are largest for $ \tau \rightarrow e $.
 To a large part this can be traced to the $ (\ln y) $--dependent terms
 in the spectrum functions as will be discussed in more detail in Sec.~6.
 On an {\sl absolute} scale the radiative corrections are generally quite
 small except for the hard end of the spectrum where they in fact
 (logarithmically) diverge \footnote{The divergent terms at the
 endpoint of the electron spectrum can be resummed into an exponential
 function \cite{kuraev85,cacciari92}.}.
 On a {\sl relative} scale the radiative corrections are quite large for
 ($ \mu \rightarrow e $) and for ($ \tau \rightarrow e $), and smaller
 for ($ \tau \rightarrow \mu $) at the soft end of the spectrum, where
 the spectrum functions are small.
 This will show up in the radiative corrections to the longitudinal
 polarization of the daughter lepton $ P_{l'}^{l} $ and the
 forward--backward asymmetry $ A_{F\!B} $ which are large in the
 threshold region for $ \mu \rightarrow e $ and $ \tau \rightarrow e $
 and small for $ \tau \rightarrow \mu $. As discussed in Sec.~3 the contribution
 of $ \beta x G_5 $ is barely discernible for the case 
 $ \tau \rightarrow \mu $ (Fig.~3a). At the scale of the figure the difference
 between the Born and NLO curves is not visible. 

 It is interesting to note that the radiative corrections go through zeros
 close to $ x = 0.68 $ and $ x = 0.82 $ for $ G_{1,3} $ and for $ G_{2,4} $,
 respectively, for all three cases discussed in this paper.
 The positions of the respective zeros are practically mass independent.
 Differences in the position of the zero show up only in the third digit.
 In fact, when discussing the $ m_{l'} \rightarrow 0 $ case in Sec.~6 we
 have checked that the positions of the zeroes remain practically unchanged
 even when letting $ y \rightarrow 0 $.
 The radiative corrections are negative and positive
 below and above the zero for $ \beta x G_{1, 4} $, resp.,
 and positive and negative below and above the zero for
 $ \beta x G_{2, 3} $.
 Qualitatively the alternating sign pattern over the range of the spectrum
 can be understood from the dominance of the $ (\ln y) $--terms and from the
 fact that the $ (\ln y) $--dependent terms have to cancel out when one
 integrates over the spectrum. 
 There is a tendency of the radiative corrections to cancel in the sums
 $ (G_1 + G_3) $ and $ (G_2 + G_4) $ and to add up in the differences
 $ (G_1 - G_3) $ and $ (G_2 - G_4) $, i.e. the radiative corrections add
 destructively and constructively in the final electron's density matrix
 elements $ \rho_{++} $ and $ \rho_{--} $, resp., to be discussed further
 in Sec.~6.  

 We next turn to the radiative corrections of the longitudinal
 polarization $ P_{e}^{l} $ of the electron and the forward--backward
 asymmetry $ A_{F\!B} $ calculated according to Eqs.~(\ref{polvec})
 and (\ref{forwardbackward1}).
 We begin by discussing the limiting value of the longitudinal
 polarization at the soft end of the spectrum including the
 radiative corrections.
 At NLO one has 
 
 \begin{Eqnarray} 
 \label{limits}
   \lim\limits_{x \rightarrow 2 y} P_e^l & = & -
   \frac{1}{3} P \cos \theta_P \Bigg\{ 1 \!-\!
   \frac{\alpha}{\pi} \frac{4 (1 - y)^2 (1 + y^2)}
   {72 y^2 \!+\! \frac{\alpha}{\pi}(5 \!-\! 10 y \!-\! 278 y^2 \!-\!
    10 y^3 \!+\! 5 y^4) \!-\! 108 \frac{\alpha}{\pi}
   \frac{1 + y}{1 - y} y^2 \ln y} \Bigg\} \qquad
   \nonumber \\[2mm] & \approx & -
   \frac{1}{3} P \cos \theta_P \Bigg\{
    1 - \frac{\alpha}{\pi} \Bigg(
    18 y^2 + \frac{5}{4} \frac{\alpha}{\pi} \Bigg)^{-1} \Bigg\}. 
 \end{Eqnarray}

 \noindent For the cases $ \mu \rightarrow e $ and
 $ \tau \rightarrow e $ the term $ (18 y^2) $ in the
 second line of Eq.~(\ref{limits}) can be neglected
 and thus the limiting value of the longitudinal
 polarization of the electron is given by
 $ P_e^l \approx  - \frac{1}{15} P \cos \theta_P $
 which is smaller than the Born term value by a
 factor of $ 5 $.
 This is evident in Figs.~2b and 4b where the radiative
 corrections to the longitudinal polarization of the electron
 are shown.
 For the $ \tau \rightarrow \mu $ case the term $ (18 y^2) $
 dominates over $ (5/4) (\alpha/\pi) $ and the limiting value
 of the longitudinal polarization of the $ \mu $ is
 $ P_e^l \approx - \frac{1}{3} P \cos \theta_P $
 $ ( 1 -  \frac{\alpha}{\pi} \frac{1}{18 y^2}) $,
 i.e. the correction to the Born term value is only
 $ \approx 3.6 \%  $.
 This can be seen in Fig.~3b.
 The NLO limiting value for the forward--backward
 asymmetry is zero since, as discussed before, a small $ \beta $--expansion shows
 that $ G_2^{(\alpha)} / G_1^{(\alpha)} $
 $ \simeq (x - x_{min})^{1/2} $ just as in the Born term case.
 This can again be seen in Figs.~2c - 4c.
 Finally, the behaviour of the transverse polarization
 $ P_e^{\perp} $ at $ x_{min} = 2 y $ is quite similar to
 that of the longitudinal polarization.
 In fact, one just has to replace
 $ \cos \theta_P \rightarrow \sin \theta_P $ and
 $ 4 \rightarrow 5 $ in the numerator of the first 
 line of Eq.(\ref{limits}) in order to obtain the
 limiting value $ P_e^{\perp} $.
 This leads to  

 \begin{equation}
   \lim\limits_{x \rightarrow 2 y}
   P_e^{\perp} \approx -
   \frac{1}{3} P \sin \theta_P \Bigg\{
   1 - \frac{\alpha}{\pi} \Bigg(
   \frac{72}{5} y^2 + \frac{\alpha}{\pi} \Bigg)^{-1} \Bigg\}. 
 \end{equation}

 \noindent Just as for $ P_e^l $ the $ O(\alpha) $ corrections are small for
 $ \tau \rightarrow \mu $ and sizeable for $ \mu \rightarrow e $ and
 $ \tau \rightarrow e $ at threshold, as can be seen in Figs.~2b - 4b. 
 Next we discuss the limiting behaviour of the spectrum functions
 at the hard end of the spectrum, where $ x \rightarrow x_{max} = 1 + y^2 $.
 As remarked on before the radiative correction contributions can be seen to
 logarithmically diverge in this limit (see footnote~7).
 Introducing $ x'_{max} = 1 + y^2 - \varepsilon $ the limiting behaviour of
 the four spectrum functions is given by

 \begin{Eqnarray} 
   \lim\limits_{x \rightarrow x_{max}} 
   \hspace{-1.5mm} G_1 & = &  - \hspace{-1.5mm}
   \lim\limits_{x \rightarrow x_{max}} 
   \hspace{-1.5mm} G_2 = - \hspace{-1.5mm}
   \lim\limits_{x \rightarrow x_{max}} 
   \hspace{-1.5mm} G_3 = \hspace{-1.5mm}
   \lim\limits_{x \rightarrow x_{max}} 
   \hspace{-1.5mm} G_4 = (1 - y^2)^2 +  
   \frac{\alpha}{\pi} 2 (1 - y^2)
   \times \\[2mm] & & \hspace{-1cm} \times
   \Bigg\{ (1 + y^2) \Bigg( \frac{3}{4} +
   \ln \Bigg( \frac{ \varepsilon}{1 - y^2} \Bigg)
   \Bigg) \ln y + (1 - y^2) \Bigg( 1 + \ln \Bigg(
   \frac{ \varepsilon}{1 - y^2} \Bigg) \Bigg)
   \Bigg\}. \nonumber
 \end{Eqnarray}

 \noindent The limiting values of the four spectrum functions are,
 up to signs, all identical.
 This implies that the radiative corrections to the longitudinal
 polarization of the daughter lepton does not change the Born term
 value $ P_{l'}^l = - 1 $ at the hard end of the spectrum irrespective
 of the values of $ P \cos \theta_P $.
 Similarly the radiative corrections to the forward--backward
 asymmetry do not change the Born term value of
 $ A_{F\!B}= - \frac{1}{2} P $ at the hard end of the spectrum.
 Finally, the NLO transverse polarization is zero at $ x = x_{max} $
 since $ G_5^{(\alpha)} $ is finite at $ x_{max} $. In fact, one has
 $ G_5^{(\alpha)} \rightarrow - (\alpha / \pi ) y (1 - y^2) \ln y $
 at $ x_{max} $.

 The fact that the relative corrections to the spectrum functions in the
 threshold region are {\sl relatively} large for ($ \mu \rightarrow e $)
 and ($ \tau \rightarrow e $) shows up in Figs.~2b and 4b, and in
 Figs.~2c and 4c, where the radiative corrections to the longitudinal
 polarization of the final state electrons and the forward--backward
 asymmetry are visibly large in the threshold region\footnote{
 In our numerical results for the longitudinal and transverse polarization
 of the daughter lepton and the forward--backward asymmetry we have
 {\sl not } expanded out the inverse of the denominator function in
 terms of powers of $ \alpha $.}.
 Note, though, that we have enhanced the threshold region in our
 presentation of $ P_{l'}^l $ and $ A_{F\!B} $ by choosing a logarithmic
 scale for $ x $ in Figs.~2b,c, 3b,c and 4b,c.
 For the forward--backward asymmetry the radiative corrections remain
 large over a larger part of the spectrum. 
 The radiative corrections to $ P_{\mu}^l $ and $ A_{F\!B} $ for the
 ($ \tau \rightarrow \mu $) decays are shown in Figs.~3b and 3c.
 As expected from the smallness of the radiative corrections
 to the spectrum functions shown in Fig.~3a the radiative
 corrections to both the longitudinal polarization of the
 final state muon and the forward--backward asymmetry are small.
 The same statement holds true for the radiative corrections
 to the transverse polarization.
 Since the spectrum functions are small at the soft end of the
 spectrum, the radiatively corrected average longitudinal polarization
 $ \langle P_{l'}^l \rangle $ of the daughter lepton $ l' $ is expected to
 remain very close to $ -1 $ in all three cases.
 Using the integrated rate functions presented at the end of this section
 we find the NLO result $ \langle P_{l'}^l \rangle = -0.999, -0.986 $ and
 $ -0.999 $ for ($ \mu \rightarrow e $), ($ \tau \rightarrow \mu $) and
 ($ \tau \rightarrow e $), resp., for $ \cos \theta_P = 0 $ with very
 little dependence on $ \cos \theta_P $.
 The corresponding figures for $ \langle A_{F\!B} \rangle $ are
 $ -0.166 $, $ -0.170 $, $ -0.166 $.
 Finally, for the transverse polarization one finds
 $ \langle P_{l'}^{\perp} \rangle = -0.0031 $,
 $ -0.037 $, $ -0.00018 $ for the three cases for $ \cos \theta_P = 0 $.

 Next we integrate the five spectrum functions
 $ \beta x G_i^{(\alpha)} $ over the electron
 spectrum according to (\ref{reducedrate}).
 One obtains

 \begin{Eqnarray} 
   \hat{G}_{1}^{(\alpha)} & = &
   \frac{\alpha}{\pi} \Bigg\{
   \frac{1}{48} (1 - y^4) (75 - 956 y^2 + 75 y^4) -
    y^4 (36 + y^4) \ln^2 y
   \nonumber \\[2mm] & - &
   \frac{\pi^2}{4} (1 - 32 y^3 + 16 y^4 - 32 y^5 + y^8) -
   \frac{1}{6} (60 + 270 y^2 - 4 y^4 + 17 y^6) y^2 \ln y
   \nonumber \\[2mm] & - &
   \frac{1}{12} (1 - y^4)(17 - 64 y^2 + 17 y^4) \ln (1 - y^2)
   \nonumber \\[5mm] & + &
    2 (1 - y)^4 (1 + 4 y + 10 y^2 + 4 y^3 + y^4) \ln (1 - y) \ln y
   \nonumber \\[5mm] & + &
    2 (1 + y)^4 (1 - 4 y + 10 y^2 - 4 y^3 + y^4) \ln (1 + y) \ln y
   \nonumber \\[5mm] & + &
   (3 + 32 y^3 + 48 y^4 + 32 y^5 + 3 y^8) \Li (-y)
   \nonumber \\[5mm] & + &
   (3 - 32 y^3 + 48 y^4 - 32 y^5 + 3 y^8) \Li ( y) \Bigg\}, \\[4mm]
   \hat{G}_{2}^{(\alpha)} & = &
   \frac{\alpha}{\pi} \Bigg\{
   \frac{1}{3} (1 - y^2) (1 + y^2 + 13 y^4 - 3 y^6)
   (\ln (1 - y) + 2 \ln (1 + y)) \ln y
   \nonumber \\[2mm] & - &
   \frac{1}{432} (1 - y)^2 (617 - 842 y + 1929 y^2 - 1592 y^3 -
    3415 y^4 + 54 y^5 - 567 y^6)
   \nonumber \\[2mm] & - &
   \frac{1}{36} (1 - y) y (12 - 18 y - 238 y^2 + 41 y^3 -
    1003 y^4 - 45 y^5 - 9 y^6) \ln y 
   \nonumber \\[2mm] & + &
   \frac{1}{36} (1 - y)^2 (13 + 26 y + 87 y^2 - 364 y^3 +
    535 y^4 - 102 y^5 - 51 y^6) \ln (1 - y)
   \nonumber \\[2mm] & + &
   \frac{1}{2} y^4 (14 + 32 y^2 - 3 y^4) \ln^2 y -
   \frac{1}{9} (1 - y^2)(25 + 13 y^2 + 67 y^4 + 15 y^6) \ln (1 + y)
   \nonumber \\[2mm] & - &
    8 y^4 \Bigg( \frac{1}{6} \ln^3 y + 
   \Bigg( 2 \Li (-y) + \Li (y) - \frac{\pi^2}{3} \Bigg) \ln y -
    6 \Ti (-y) - \Ti (y) - \frac{7}{2} \zeta (3) \Bigg)
   \nonumber \\[2mm] & + &
   \frac{1}{3} (7 + 24 y^2 + 48 y^4 - 8 y^6 + 9 y^8)
   \Bigg( \Li (-y) + \frac{\pi^2}{12} \Bigg) \Bigg\}, \\[4mm]
   \hat{G}_{3}^{(\alpha)} & = &
   \frac{\alpha}{\pi} \Bigg\{
   \frac{1}{3} (1 - y^2) (3 - 13 y^2 - y^4 - y^6)
   (\ln (1 - y) + 2 \ln (1 + y)) \ln y
   \nonumber \\[2mm] & - &
   \frac{1}{432} (1 - y)^2 (567 - 54 y + 3415 y^2 + 1592 y^3 -
    1929 y^4 + 842 y^5 - 617 y^6)
   \nonumber \\[2mm] & - &
   \frac{1}{36} (1 - y) y (36 + 98 y + 590 y^2 - 265 y^3 +
    27 y^4 - 99 y^5 - 87 y^6) \ln y
   \nonumber \\[2mm] & + &
   \frac{1}{36} (1 - y)^2 (51 + 102 y - 535 y^2 + 364 y^3 -
    87 y^4 - 26 y^5 - 13 y^6) \ln (1 - y)
   \nonumber \\[2mm] & - &
   \frac{1}{6} y^2 (8 + 66 y^2 - 24 y^4 - y^6) \ln^2 y -
   \frac{1}{9} (1 \!-\! y^2) (15 + 67 y^2 + 13 y^4 + 25 y^6) \ln (1 + y)
   \nonumber \\ & + & 8 y^4 \Bigg( \frac{1}{6} \ln^3 y + 
   \Bigg( 2 \Li (-y) + \Li (y) - \frac{\pi^2}{3} \Bigg) \ln y -
    6 \Ti (-y) - \Ti (y) - \frac{7}{2} \zeta (3) \Bigg)
   \nonumber \\[2mm] & + &
   \frac{1}{3} (9 - 8 y^2 + 48 y^4 + 24 y^6 + 7 y^8)
   \Bigg( \Li (-y) + \frac{\pi^2}{12} \Bigg) \Bigg\}, \\[4mm]
   \hat{G}_{4}^{(\alpha)} & = &
   \frac{\alpha}{\pi} \Bigg\{
   \frac{1}{432} (1 - y^4) (581 + 6140 y^2 + 581 y^4)
   \nonumber \\[2mm] & - &
   \frac{\pi^2}{36} (7 - 6 y + 28 y^2 + 222 y^3 - 164 y^4 +
    390 y^5 - 140 y^6 + 66 y^7 - 5 y^8)
   \nonumber \\[2mm] & - &
   \frac{1}{36} (1 - y^4) (13 - 176 y^2 + 13 y^4) \ln (1 - y^2) +
   \frac{1}{3} y^2 (2 + 46 y^2 + 38 y^4 + y^6) \ln^2 y
   \nonumber \\[2mm] & + &
   \frac{2}{3} y (1 - y)^4 (5 - 4 y + 5 y^2)
   \ln (1 - y) \ln y -
   \frac{2}{3} y (1 + y)^4 (5 + 4 y + 5 y^2)
   \ln (1 + y) \ln y
   \nonumber \\[2mm] & + &
   \frac{1}{18} y^2 (212 + 930 y^2 + 388 y^4 - 13 y^6) \ln y
   \nonumber \\[2mm] & + &
   \frac{1}{3} (1 - 10 y - 56 y^2 - 102 y^3 - 164 y^4 -
    102 y^5 - 56 y^6 - 10 y^7 + y^8) \Li (-y)
   \nonumber \\[2mm] & + &
   \frac{1}{3} (1 + 10 y - 56 y^2 + 102 y^3 - 164 y^4 +
    102 y^5 - 56 y^6 + 10 y^7 + y^8) \Li (y) \Bigg\}, \\[4mm]
   \hat{G}_{5}^{(\alpha)} & = &
   \frac{\alpha}{\pi} y \Bigg\{ -
   \frac{7}{18} (1 - y^2)(5 + 34 y^2 + 5 y^4)
   \nonumber \\[2mm] & - &
   \frac{2}{3} y^2 (9 + 18 y^2 - y^4) \ln^2 y +
   \frac{\pi^2}{6} (1 - 3 y^2 + 32 y^3 - 3 y^4 + y^6)
   \nonumber \\[2mm] & - &
   \frac{4}{3} (1 - y)^4 (1 + 4 y + y^2) \ln (1 - y) \ln y -
   \frac{4}{3} (1 + y)^4 (1 - 4 y + y^2) \ln (1 + y) \ln y
   \nonumber \\[2mm] & - &
   \frac{1}{18} (15 + 249 y^2 + 141 y^4 - 29 y^6) \ln y +
   \frac{1}{9} (1 - y^2) (11 + 38 y^2 + 11 y^4) \ln (1 - y^2)
   \nonumber \\[2mm] & - &
   \frac{2}{3} (3 - 9 y^2 - 32 y^3 - 9 y^4 + 3 y^6) \Li (-y)
   \nonumber \\[2mm] & - &
   \frac{2}{3} (3 - 9 y^2 + 32 y^3 - 9 y^4 + 3 y^6) \Li (y) \Bigg\}.
 \end{Eqnarray}
 
 \noindent Note that $ \hat{G}_2^{(\alpha)} $ and
 $ \hat{G}_3^{(\alpha)} $ contain trilog functions,
 and associated with them, Euler's $ Zeta $ function
 $ \zeta(3) $, whereas $ \hat{G}_1^{(\alpha)} $,
 $ \hat{G}_4^{(\alpha)} $ and $ \hat{G}_5^{(\alpha)} $
 contain only dilog functions.
 In agreement with the Lee--Nauenberg theorem \cite{lee64}
 the rate functions do not contain any logarithmic mass singularities.
 Our result for $ \hat{G}_1^{(\alpha)} $ agrees with the result
 in \cite{nir89} where a different route of phase space integrations
 was taken to arrive at the total rate.
 Our result for $ \hat{G}_2^{(\alpha)} $ agrees with the result
 in \cite{arbuzov01}.
 The results on $ \hat{G}_3^{(\alpha)} $, $ \hat{G}_4^{(\alpha)} $
 and $ \hat{G}_5^{(\alpha)} $ are new.
 As an additional check we have checked that all
 five rate functions $ \hat{G}_i^{(\alpha)} $
 vanish for $ y \rightarrow 1 $.

 \renewcommand{\arraystretch}{1.5}
 \renewcommand{\tabcolsep}{0mm}
 \begin{table}
 \rule{160mm}{1pt}
 \begin{tabular*}{160mm}{l@{\extracolsep{\fill}}|lll|lll|lll} 
  \hline
  & \multicolumn{3}{c|}{$ \mu  \rightarrow  e  $} 
  & \multicolumn{3}{c|}{$ \tau \rightarrow \mu $}
  & \multicolumn{3}{c}{$ \tau \rightarrow  e  $} \\
  $ i $
  & $ \delta \hat{G}_i^{<} $ & $ \delta \hat{G}_i^{>} $ & $ \delta \hat{G}_i $
  & $ \delta \hat{G}_i^{<} $ & $ \delta \hat{G}_i^{>} $ & $ \delta \hat{G}_i $
  & $ \delta \hat{G}_i^{<} $ & $ \delta \hat{G}_i^{>} $ & $ \delta \hat{G}_i $
  \\ \hline
   $ 1 \, $ &
   $ + 2.80 \% $ & $ - 2.69 \% $ & $ - 0.42 \% \, $ &
   $ + 0.76 \% $ & $ - 1.14 \% $ & $ - 0.40 \% \, $ &
   $ + \zz 5.20 \% $ & $ - 4.44 \% $ & $ - 0.42 \% \, $ \\
   $ 2 \, $ &
   $ + 8.22 \% $ & $ - 3.70 \% $ & $ - 0.68 \% \, $ &
   $ + 2.51 \% $ & $ - 1.59 \% $ & $ - 0.62 \% \, $ &
   $ + 14.70 \% $ & $ - 6.10 \% $ & $ - 0.68 \% $\\
   $ 3 \, $ &
   $ + 2.56 \% $ & $ - 2.68 \% $ & $ - 0.53 \% \, $ &
   $ + 0.66 \% $ & $ - 1.14 \% $ & $ - 0.45 \% \, $ &
   $ + \zz 4.95 \% $ & $ - 4.43 \% $ & $ - 0.54 \% $ \\
   $ 4 \, $ &
   $ + 7.82 \% $ & $ - 3.70 \% $ & $ - 0.79 \% \, $ &
   $ + 2.53 \% $ & $ - 1.59 \% $ & $ - 0.67 \% \, $ &
   $ + 14.28 \% $ & $ - 6.10 \% $ & $ - 0.80 \% $ \\
   $ 5 \, $ &
   $ + 0.76 \% $ & $ - 3.68 \% $ & $ - 2.89 \% \, $ &
   $ + 0.09 \% $ & $ - 1.62 \% $ & $ - 1.49 \% \, $ &
   $ + \zz 1.72 \% $ & $ - 6.03 \% $ & $ - 4.53 \% $ \\
  \hline
 \end{tabular*}
 \rule{160mm}{1pt} \\[0.8ex]
 \centerline{\bf Table 2:} 
 {\small Numerical values of partially integrated
  ($ \delta \hat{G}_i^{<}, \, \delta \hat{G}_i^{>} $)
  and total rate functions ($ \delta \hat{G}_i $)
  divided by their respective Born term values.
  The symbols ``$ < $'' and ``$ > $'' stand for
  integrations from threshold to the zero point of the
  respective $ O(\alpha) $ contributions, and from the
  zero point to the endpoint of the spectrum.}
 \end{table}

 In order to get a quantitative feeling about the size of the radiative
 corrections to the respective spectrum functions we have listed in Tab.~2
 the percentage changes $ \delta \hat{G_i} $ induced by the radiative
 corrections, where

 \begin{equation} 
   \label{percentage}
   \delta \hat{G_i} =
   \frac{(\hat{G}_i^{(\alpha)} + \hat{G_i}^{Born}) - \hat{G_i}^{Born}}
   {\hat{G_i}^{Born}} =
   \frac{\hat{G}_i^{(\alpha)}}{\hat{G_i}^{Born}}.
 \end{equation}

 \noindent The relative radiative corrections $ \delta \hat{G_i} $ can
 all be seen to be close to the naive expectation of $ O(\alpha) $ where
 the relative radiative corrections to $ \hat{G}_5 $ are largest.
 However, as was emphasized earlier on, all radiative corrections
 discussed in this paper go through zeros.
 Integrating over the whole spectrum therefore does not give an
 adequate representation of the size of the radiative corrections
 to the spectrum since there are sizable cancellation effects.
 This cancellation would become less effective if moments of
 the spectrum functions were taken.
 The moments could be chosen such that they either
 emphasize the threshold or the endpoint region.
 Alternatively, one can consider partially integrated rates where the
 integrations either run from threshold to the point where the radiative
 corrections go to zero or from the zero point to the endpoint.
 The two partially integrated rate functions will be denoted by
 $ \hat{G}_i^{<} $ (lower part) and by $ \hat{G}_i^{>} $ (upper part).
 The two respective partially integrated relative rate functions
 $ \delta \hat{G}_i^{<} $ and $ \delta \hat{G}_i^{>} $ are also listed
 in Tab.~2.
 The relative radiative corrections for the partially integrated rate
 functions can be seen to be much larger than for the fully integrated
 rate functions and can amount up to $ O(10 \%) $ which is much larger
 than the naive $ O(\alpha) $ expectation.
 Note, though, that, in contrast to the rate functions, neither the
 moments of the spectral functions nor the partially integrated rates
 are free of logarithmic mass singularities. 

 Since the complete rate expressions are rather unwieldy it is useful to
 consider the small $ y $--expansions of the rate expressions. One has

 \begin{Eqnarray} 
   \hat{G}_{1}^{(\alpha)} & = & \frac{\alpha}{\pi} \Bigg\{
   \frac{25 - 4 \pi^2}{16} - (17 + 12 \ln y) y^2 + 8 \pi^2 y^3 +
    O(y^4) \Bigg\}, \\[2mm]
   \hat{G}_{2}^{(\alpha)} & = & \frac{\alpha}{\pi} \Bigg\{ -
   \frac{617 - 84 \pi^2}{432} - \frac{2}{3} y -
   \frac{1}{3} (24 - 2 \pi^2 - \ln y) y^2 + \nonumber \\[2mm] & + &
   \frac{2}{27} (71 + 84 \ln y) y^3 + O(y^4) \Bigg\}, \\[2mm]
   \hat{G}_{3}^{(\alpha)} & = & \frac{\alpha}{\pi} \Bigg\{ -
   \frac{21 - 4 \pi^2}{16} - \frac{10}{3} y -
   \frac{1}{27} (232 + 6 \pi^2 + 87 \ln y + \nonumber \\[2mm] & + &
    36 \ln^2 y) y^2 + \frac{2}{9} (121 - 84 \ln y) y^3 +
    O(y^4) \Bigg\}, \\[2mm]
   \hat{G}_{4}^{(\alpha)} & = & \frac{\alpha}{\pi} \Bigg\{
   \frac{7 (83 - 12 \pi^2)}{432} + \frac{1}{6} y +
   \frac{1}{27} (578 - 21 \pi^2 + 138 \ln y + \nonumber \\[2mm] & + &
    18 \ln^2 y) y^2 - \frac{37}{6} \pi^2 y^3 + O(y^4) \Bigg\}, \\[2mm]
   \hat{G}_{5}^{(\alpha)} & = & \frac{\alpha}{\pi} \,y\, \Bigg\{ -
   \frac{35 - 3 \pi^2 + 15 \ln y}{18} -
   \frac{1}{2} (27 + \pi^2 + 25 \ln y + \nonumber \\[2mm] & + &
   12 \ln^2 y ) y^2 + \frac{16}{3} \pi^2 y^3 + O(y^4) \Bigg\}.
 \end{Eqnarray}

 It is well--known that the $ O(\alpha) $ small--$ y $ corrections to
 the reduced rate $ \hat{G}_1^{(\alpha)} $ start only at $ O(y^2) $
 (see e.g. \cite{roos71,vanrit00}).
 In contradistinction and contrary to the Born term case
 the mass corrections to the spin dependent rate functions
 $ \hat{G}_2^{(\alpha)} $, $ \hat{G}_3^{(\alpha)} $ and
 $ \hat{G}_4^{(\alpha)} $ all start at $ O(y) $.

 In order to obtain a quantitative feeling about the
 importance of mass effects in the $ O(\alpha) $ radiative
 contributions we have listed in Tab.~1 the percentage
 changes in the rate functions $\hat{G}_i $ when going
 from $ m_{l'} \rightarrow 0 $ to $ m_{l'} \ne 0 $
 using the $ m_{l'} \rightarrow 0 $ results listed in Sec.~6.
 The mass effects are not small, in particular for the case
 $ (\tau \rightarrow \mu) $.
 The percentage changes for the radiative contributions are
 larger than those in the Born term case which is partly due
 to the difference in the power pattern of the final state
 lepton mass corrections in the two cases.
 The quality of the $ m_{l'} \rightarrow 0 $ approximation
 for the radiative corrections can be assessed by referring
 again to Tab.~1.
 The final state lepton mass effects tend to reduce the
 overall size of the radiative corrections.
 The reduction is largest $ (O(10 \%)) $  for the case 
 $ (\tau \rightarrow \mu) $ and smallest $ (O(10^{-1})) $
 for the case $ (\tau \rightarrow e) $.
 They are largest for the rate functions $ \hat{G}_3 $
 and $ \hat{G}_4 $ and smallest for $ \hat{G}_1 $.
 That the mass effects are smallest for $ \hat{G}_1 $
 is due to the fact that the mass corrections to
 $ \hat{G}_1 $ set in only at $ O(y^2) $ (see Eq.~(47)).
 Whether one is willing to tolerate the error incurred in
 using the simpler $ m_{l'} \rightarrow 0 $ radiative correction
 formulas depends of course on the accuracy required for the
 application at hand.

 We now turn to the discussion of the NLO average
 longitudinal and transverse polarization of the electron
 $\langle P_e^l \rangle $ and $\langle P_e^{\perp} \rangle $,
 and the NLO average forward--backward asymmetry
 $ \langle A_{F\!B} \rangle $.
 They take remarkably simple forms in the $ y \rightarrow 0 $ limit.
 Including the Born term contribution and expanding the inverse 
 denominator in powers of $ {\alpha} / {\pi} $ one has at NLO

 \begin{equation} 
   \label{longpolaverage}
   \lim\limits_{y \rightarrow 0}
   \langle P_{e}^l \rangle = - (1 - \frac{\alpha}{2 \pi}).
 \end{equation}

 \noindent Note that $ \langle P_e^l \rangle $ does not
 depend on $ P \cos \theta_P $ in this approximation.

 \noindent For the forward--backward asymmetry one obtains
 
 \begin{equation} 
   \label{fbaverage1}
   \lim\limits_{y \rightarrow 0}
   \langle A_{F\!B} \rangle = -
   \frac{1}{6} P (1 - \frac{\alpha}{\pi} \,
   \frac{6 \pi^2 - 49}{9}).
 \end{equation}
 
 \noindent Finally, for the transverse polarization one obtains in 
 the same approximation ($\cos \theta = 0 $)

 \begin{equation} 
   \label{transpolaverage}
   \lim\limits_{y \rightarrow 0}
   \langle P^{\perp}_e \rangle = -
   \frac{2}{3} \, P \, y (1 + \frac{\alpha}{\pi } \,
   \frac{65 + 60 \ln y }{24}).
 \end{equation}
 
 \noindent The $ O(\alpha) $ corrections to $ \langle P_e^l \rangle $,
 $ \langle A_{F\!B} \rangle $ and to $ \langle P^{\perp}_e \rangle $
 are thus quite small.
 The actual numerical values for $ \langle P_e^l \rangle $,
 $ \langle A_{FB} \rangle $ and for $ \langle P^{\perp}_e \rangle $ listed 
 earlier in this section lie very close to the above estimates in all three
 cases. 
   
 
\section{The \protect \boldmath $ m_e \rightarrow 0 $ limit
         and the anomalous helicity flip contribution}

 The purpose of this section is two-fold.
 First we discuss the $ m_e \rightarrow 0 $ limit of the
 $ m_e \ne 0 $ $ O(\alpha) $ results given in Sec.~5.
 This allows us to make contact with the $ m_e \rightarrow 0 $
 results derived previously \cite{fs74}.
 Second we discuss in some detail the origin of the anomalous
 helicity flip contribution resulting from collinear
 photon emission of the electron.
 Our results are presented in terms of the two
 diagonal components of the density matrix of
 the final state electron, which, in the limit
 $ m_e \rightarrow 0 $, are nothing but the helicity
 no--flip and helicity flip contributions
 of the final state electron.
 The naive prediction of massless QED is that the helicity flip
 contributions vanish in all orders of perturbation theory.
 However, as first pointed out by Lee and Nauenberg \cite{lee64},
 there will be a nonzero helicity flip contribution from collinear
 photon emission which survives the $ m_e \rightarrow 0 $ limit.
 This will be demonstrated in our $ O(\alpha) $
 $ m_e \rightarrow 0 $ expressions.
 Our $ m_e \rightarrow 0 $ result for the helicity flip contribution
 is found to be in agreement with expectations derived from the
 universal equivalent particle approach of Falk and Sehgal \cite{falk94}.
 For a discussion of the quality of the  $ m_{l'} \rightarrow 0 $
 approximation for the three cases $ (\mu \rightarrow e) $,
 $ (\tau \rightarrow \mu) $ and $ (\tau \rightarrow e) $ we
 refer to the discussion at the end of Sec.~5.  

 Since we want to discuss the helicity no--flip and
 flip contributions separately it is convenient to
 choose a slightly different representation for the
 differential rate Eq.~(\ref{diffrate}).
 We write the differential rate in terms
 of the no--flip and flip contributions

 \begin{equation} 
   \label{flip-noflip1}
    \frac{d \Gamma}{dx \, d\!\cos \theta_P} =
    \frac{d \Gamma^{n\!f}}{dx \, d\!\cos \theta_P} (1 - \cos \theta) +
    \frac{d \Gamma^{h\!f}}{dx \, d\!\cos \theta_P} (1 + \cos \theta),
 \end{equation}

 \noindent where the no--flip and flip contributions are given by

 \begin{equation} 
   \label{flip-noflip2}
    \frac{d \Gamma^{n\!f/h\!f}}{dx \, d\!\cos \theta_P } =
    \frac{1}{2} \beta x \, \Gamma_0
    \Big( ( G_1 \mp G_3) + ( G_2 \mp G_4) P \cos \theta_P \Big). 
 \end{equation}

 \noindent The terminology helicity no--flip $ (n\!f) $ and helicity flip
 $ (h\!f) $ is really only appropriate in the $ m_e \rightarrow 0 $ limit
 where the electron emerging from the left--chiral weak interaction current
 is purely left--handed.
 After photon emission the electron can then remain
 left--handed $ (n\!f) $ or can become right--handed $ (h\!f) $.
 For $ m_e \ne 0 $ the respective $ n\!f $ and $ h\!f $ contributions
 are nothing but the (unnormalized) diagonal elements of the density
 matrix of the electron, i.e. $ \Gamma^{n\!f} \sim \rho_{--} $ and
 $ \Gamma^{h\!f} \sim \rho_{++} $. 
 
 As concerns the helicity flip contribution one notes that there
 are no Born term helicity flip contributions in the limit $ m_e = 0 $
 since a massles electron emerging from the weak $ (V-A) $
 vertex is left--handed.
 This is explicitly seen by inserting the Born term rate functions
 Eq.~(\ref{bornratefunctions}) in Eq.~(\ref{flip-noflip2}).
 Naively, one would expect no helicity flip
 contributions also at $ O(\alpha) $ because,
 in massless QED with $ m_e = 0 $, photon emission
 from the electron is helicity conserving.
 However, taking the limit $ m_e \rightarrow 0 $ in
 Eqs.~(\ref{spectrum1}) - (\ref{spectrum4}) one
 finds helicity flip contributions which survive
 the $ m_e \rightarrow 0 $ limit.
 In fact, one finds 

 \begin{equation}
   \label{flip}
   \frac{d \Gamma^{h\!f}}{dx \, d\!\cos \theta_P } =
   \frac{\alpha}{12 \pi} \Gamma_0
   \Big( \Big[ (1 - x)^2 (5 - 2 x) \Big] -
   \Big[ (1 - x)^2 (1 + 2 x) \Big] P \cos \theta_P \Big),
 \end{equation}

 \noindent which agrees with the result presented in Ref.~\cite{fs74}.
 Because of the naive expectation that the helicity flip contribution
 vanishes in massless QED the presence of a helicity flip contribution
 is sometimes referred to as the anomalous helicity flip contribution.
 Moreover, the authors of Ref.~\cite{dz71} were able to show that the
 well--known axial anomaly can be traced to the existence of an anomalous
 helicity flip contribution to the absorptive part of the $ VV\!A $
 triangle diagram in massless QED.
 This gives further justification for the use of the terminology
 ``anomalous helicity flip contribution''.
 In order to set the anomalous contributions apart we have
 highlighted them in Eq.~(\ref{flip}) by enclosing them in
 square brackets.

 The authors of Ref.~\cite{fs74} had already expressed surprise
 at the simplicity of the structure of the $ O(\alpha) $
 helicity flip contributions without, however, attempting to
 identify the source of this structural simplicity.
 The simplicity of the helicity flip contribution becomes manifest
 in the equivalent particle description of $ \mu $--decay where,
 in the peaking approximation, $ \mu $--decay is described by the
 two--stage process $ \mu^{-} \rightarrow e^{-} $ followed by the
 branching process $ e^{-} \rightarrow e^{-} + \gamma $ characterized
 by universal splitting functions $ D_{n\!f/h\!f}(z) $ \cite{falk94}.
 In the splitting process $ z $ is the fractional energy of the
 emitted photon \cite{falk94}.
 The off--shell electron in the propagator is replaced by an
 equivalent on--shell electron in the intermediate state.
 Since the helicity flip contribution arises entirely
 from the collinear configuration it can be calculated
 in its entirety using the equivalent particle description.

 The helicity flip splitting function is given by
 $ D_{h \! f}(z) = \alpha z/(2\pi) $, where
 $ z = k_0 / E' = (E' - E)/ E' = 1 - x/x' $, and where $ k_0 $ is the
 energy of the emitted photon \cite{falk94}.
 $ E' $ and $ E $ denote the energies of the initial and final
 electron in the splitting process.
 The helicity flip splitting function has to be folded with the
 appropriate $ m_e = 0 $ Born term contribution.
 The lower limit of the folding integration is determined
 by the soft photon point where $ E' = E $.
 The upper limit is determined by the maximal energy of the
 initial electron $ E' = m_{\mu}/{2} $. 
 One obtains 

 \begin{Eqnarray} 
   \label{anomalous}
   \frac{d \Gamma^{h\!f}}{dx \, d\!\cos \theta_P } & = &
   \frac{\alpha}{2 \pi}
   \int_x^1 dx' \frac{1}{x'}
   \frac{d \Gamma^{Born; n\!f} (x')}
    {dx' \, d\!\cos \theta_P }
    (1 - \frac{x}{x'}) \nonumber \\[3mm] & = &
   \frac{\alpha}{2 \pi} \Gamma_0
   \int_x^1 dx' (x' - x)
   \Big( (3 - 2 x') + (1 - 2 x')
    P \cos \theta_P \Big) \nonumber \\[3mm] & = &
   \frac{\alpha}{12 \pi} \Gamma_0
   \Big( (1 - x)^2 (5 - 2 x) -(1 - x)^2 (1 + 2 x)
    P \cos \theta_P \Big),
 \end{Eqnarray}

 \noindent which exactly reproduces
 the result of Eq.~(\ref{flip}).
 Note that the flip spectrum function does not
 contain a logarithmic mass factor.
 Integrating over the spectrum one obtains

 \begin{equation} 
   \label{intspechf}
   \frac{d \langle \Gamma^{h\!f} \rangle}{d\!\cos \theta_P } =
   \frac{\alpha}{\pi} \Gamma_0 \Bigg(
   \Bigg[ \frac{1}{8} \Bigg] -
   \Bigg[ \frac{1}{24} \Bigg] P \cos \theta_P \Bigg).
 \end{equation}

 \noindent which can be checked to agree with the
 results in Sec.~5 setting $ y = 0 $.

 The helicity no--flip contribution is again obtained by taking the
 $ m_e \rightarrow 0 $ limit in Eqs.~(\ref{spectrum1}) - (\ref{spectrum4})
 but now for the differences of the respective spectrum functions as
 specified in Eq.~(\ref{flip-noflip2}).
 Including the $ O(y^0) $ Born term contributions one obtains

 \begin{Eqnarray}
   \label{no--flip}
   \frac{d \Gamma^{n\!f}}{dx \, d\!\cos \theta_P } & = &
   \Gamma_0 \Bigg( x^2(3 - 2 x) + \frac{\alpha}{12 \pi}
   \Bigg\{ \!\! - \Bigg[ (1 - x)^2 (5 - 2 x) \Bigg] -
    4 x (11 - 10 x + 5 x^2) \nonumber \\[3mm] & - &
    6 (6 - x) x \ln (1 - x) - 2 (5 + 12 x - 15 x^2 + 4 x^3)
   \ln \Bigg( \frac{y}{x} \Bigg) \nonumber \\[3mm] & + &
    6 (3 - 4 x) x^2 \ln \Bigg( \frac{y}{1 - x} \Bigg) +
    12 (3 - 2 x) x^2 \sigma \Bigg\} \nonumber \\[3mm] & + &
   \Bigg( x^2(1-2x) + \frac{\alpha}{12 \pi}
   \Bigg\{ \Bigg[ (1 - x)^2 (1 + 2 x) \Bigg] -
    2 (3 - 10 x - 13 x^2 + 8 x^3) \nonumber \\[3mm] & - &
   \frac{2}{x} (4 - 12 x + 18 x^2 - 13 x^3)
   \ln (1 - x) + 2 (1 + 21 x^2 - 4 x^3)
   \ln \Bigg( \frac{y}{x} \Bigg) \nonumber \\[3mm] & + &
    6 (1 - 4 x) x^2 \ln \Bigg( \frac{y}{1 - x} \Bigg) +
    12 (1 - 2 x) x^2 \sigma \Bigg\} \Bigg)
    P \cos \theta_P \Bigg),
  \end{Eqnarray}

 \noindent where

 \begin{equation} 
   \sigma = - \frac{\pi^2}{3} +
    2 \Li(x) + \ln(x) \ln(1 - x) +
    2 \ln \Bigg( \frac{x}{1 - x} \Bigg)
   \ln \Bigg( \frac{y}{x} \Bigg).
 \end{equation}

 \noindent The no--flip contribution agrees with the result presented in
 Ref.~\cite{fs74}. We have highlighted the anamolous contributions in 
 Eq.~(\ref{no--flip}) by enclosing them in square brackets.
 When calculating the total spectrum, i.e. when summing the flip and no--flip
 contributions (\ref{flip}) and (\ref{no--flip}), the anomalous contributions
 cancel.

 The $ O(\alpha) $ no--flip contribution is much larger
 than the flip contribution.
 This is illustrated in Fig.~5 for the electron spectrum
 in the  $ \mu \rightarrow e $ decay where the flip contribution
 has been multiplied by a factor of 20 in order to make it visible at all.
 The vanishing of the flip contribution for $ \cos \theta = + 1 $
 at the hard end of the spectrum can be understood from
 angular momentum conservation.
 In order to be able to discuss the residual mass dependence
 the no--flip contribution has been split into its constant part
 and its logarithmic $ (\ln y) $ part which come in with opposite
 signs over most of the spectrum, i.e. they partially cancel
 in the spectrum.
 Considering the numerical values of the ratios
 $ \ln y_{(\tau \rightarrow \mu)} / \ln y_{(\mu \rightarrow e )} = 0.53 $ and
 $ \ln y_{(\tau \rightarrow e)}) / \ln y_{(\mu \rightarrow e )} = 1.53 $ it is
 clear from Fig.~5 that the cancellation between the constant and the
 logarithmic part is strongest for $ \tau \rightarrow \mu $ and weakest for
 $ \tau \rightarrow e $.
 This observation provides a qualitative explanation of the
 hierarchy of the size of radiative corrections to the three decay cases
 as described in Sec.~5, namely the radiative corrections are largest for
 $ \tau \rightarrow e $ and smallest for $ \tau \rightarrow \mu $.

 We now turn to the discussion of the longitudinal polarization
 of the daughter lepton in the $ m_e \rightarrow 0 $ limit.
 It takes a rather simple form when one expands out
 the inverse denominator in terms of powers of $ (\alpha / \pi) $
 keeping only the $ O(\alpha) $ contribution.
 In terms of the flip and no--flip contributions
 the longitudinal polarization of the electron reads
 
 \begin{equation}
    P_{e}^l = \frac
    {d\Gamma^{hf} - d\Gamma^{nf}}
    {d\Gamma^{hf} + d\Gamma^{nf}}
 \end{equation}

 \noindent When expanding the inverse
 denominator in terms of powers of
 $ (\alpha / \pi) $ one sees that the
 ``normal" contributions in the numerator
 and denominator cancel exactly at $ O(\alpha) $
 and one just remains with the anomalous contributions
 (No such cancellations occur for the
  forward-backward asymmetry $ A_{FB} $.
  We therefore refrain from presenting a
  closed formula for $ A_{FB} $ in this
  approximation).
 One obtains

 \begin{equation}  
   \label{longpol3}
    P_{e}^l = - \Bigg( 1 - \frac{\alpha}{6\pi} \;
    \frac{(1 - x)^2}{x^2} \;
    \frac{5 - 2 x - (1 + 2 x) P \cos \theta_P}
    {3 - 2 x + (1 - 2 x) P \cos \theta_P} \Bigg)
 \end{equation}
 
 \noindent It is clear that Eq.~(\ref{longpol3})
 does not apply very close to threshold.
 Due to the factor $ (1 - x)^2 / x^2 $ in (\ref{longpol3})
 the radiative corrections to the longitudinal polarization
 of the daughter lepton are largest close to threshold as
 is evidenced in the plots Fig.~2b and Fig.~4b.
 In the case of $ (\tau \rightarrow \mu) $, mass effects
 prevent the radiative corrections to become large in
 the threshold region.

 Integrating the no--flip contribution over the spectrum one obtains

 \begin{equation} 
   \label{intspecnf}
   \frac{d \langle \Gamma^{n\!f} \rangle}{d\!\cos \theta_P } =
   \Gamma_0 \Bigg( \frac{1}{2} +
   \frac{\alpha}{\pi} \Bigg( \!\! -
   \Bigg[ \frac{1}{8} \Bigg] + \frac{25}{16} -
   \frac{1}{4} \pi^2 \Bigg) +
   \Bigg( \!\! - \frac{1}{6} +
   \frac{\alpha}{\pi} \Bigg(
   \Bigg[ \frac{1}{24} \Bigg] -
   \frac{617 - 84 \pi^2}{432} \Bigg) \Bigg)
    P \cos \theta_P \Bigg).
 \end{equation}
 
 \noindent which agrees with the results in Sec.~5.
 In particular one reproduces the simple expression
 Eq.(\ref{longpolaverage}) for the average of the longitudinal polarization.
 We have again highlighted the anomalous contributions
 by enclosing them in square brackets.
 The anomalous contributions can be seen to cancel in
 the spectrum and rate functions when adding up the
 spectrum no--flip contribution (\ref{no--flip})
 and flip contribution (\ref{flip}), and the respective
 rate contributions (\ref{intspecnf}) and (\ref{intspechf}).
 Numerically the $ O(\alpha) $ no--flip
 contribution dominates over the $ O(\alpha) $
 (anomalous) flip contribution.
 For the $ O(\alpha) $ contributions to the
 rate functions, which do not depend on $ y $ 
 in the $ y \rightarrow 0 $ approximation,
 one obtains ($ -0.168 $, $ -0.121 $, $ -0.107 $)
 for the rate ratio $ d\Gamma^{h\!f(\alpha)} /d\Gamma^{n\!f (\alpha)} $,
 for $ \cos \theta_P = 1, 0, -1 $.


\section{Summary and conclusions}

 We have computed the $ O(\alpha) $ corrections to the leptonic decays
 of the $ \mu $-- and $ \tau $--leptons including polarization effects
 and the full mass dependence of the respective final--state leptons.
 The radiative corrections to the spectrum functions are sizeable for
 $ (\mu \rightarrow e) $--decays, large for $ (\tau \rightarrow e) $--decays,
 and smaller for $ (\tau \rightarrow \mu) $--decays.
 In large part this pattern is due to the
 $ (\ln y) $--dependent contributions to the spectrum.
 The polarization of the final-state lepton deviates
 substantially from the naive $ m_{l'} = 0 $ values $ P^{l}_{l'} = -1 $
 and $ P^\perp_{l'} = 0 $ towards the soft end of the spectrum.
 The radiative corrections to the longitudinal and transverse polarization
 of the daughter lepton in the threshold region are substantial for
 $ (\mu \rightarrow e) $-- and
 $ (\tau \rightarrow e) $--decays and small for
 $ (\tau \rightarrow \mu) $--decays.
 Similar statements hold for the 
 forward--backward asymmetry $ A_{FB} $.

 For the rate functions we have compared our $ O(\alpha) $ $ m_{l'} \ne 0 $
 results with $ m_{l'} \rightarrow 0 $ results derived previously in
 \cite{fs74}. In particular in the $ (\tau \rightarrow \mu) $ case the errors
 incurred in using the  $ O(\alpha) $ $ m_{l'} \rightarrow 0 $ results are
 large (of $ O(10\%) $ in the $ O(\alpha) $ rate functions). A mass effect
 is already showing up in the experimental values for the branching ratios of
 the two decay modes
 $ \tau^{-} \rightarrow \mu^{-} + \bar{\nu}_{\mu} + \nu_{\tau} $ and 
 $ \tau^{-} \rightarrow e^{-} + \bar{\nu}_{e} + \nu_{\tau} $. They are
 $ BR (\tau^{-} \rightarrow \mu^{-} + \bar{\nu}_{\mu} + \nu_{\tau})=
 (17.37 \pm 0.06) \% $ and 
 $ BR (\tau^{-} \rightarrow e^{-} + \bar{\nu}_{e} + \nu_{\tau}) =
 (17.84 \pm 0.06) \% $ \cite{hagiwara02}. The two branching ratios are
 compatible with the mass dependence of the Born term rates. In order to
 be sensitive to the mass dependence of the radiative corrections the error
 on the branching ratios would have to be improved by at least a factor of ten.

 Whether one is willing to tolerate the error brought about by
 using the simpler $ m_{l'} \rightarrow 0 $ radiative correction
 formulas depends of course on the accuracy required for the
 application at hand.
 We nevertheless strongly recommend use of the complete
 results in numerical investigations instead of using the
 $ m_{l'} \rightarrow 0 $ approximation.
 The analytical $ m_{l'} \ne 0 $ formulas written down in this
 paper are of sufficient simplicity to allow for easy incorporation
 into numerical programs.

 From what was being said in Sec.3 it is clear that the results
 of this paper can immediately be applied to the case of semileptonic
 quark decays, where e.g. in the case of the semileptonic
 $ b \rightarrow c + l^{-} + \bar{\nu}_{l} $ decays
 the final state $ c $--quark mass can certainly not be neglected.

 \vspace{1cm} {\bf Acknowledgements:}
 We acknowledge informative discussions with F.~Scheck, K. Schilcher and
 H.~Spiesberger. We would like to thank A.B.~Arbuzov for clarifying e--mail
 exchanges. M.C.~Mauser and S.~Groote are supported by the DFG (Germany)
 through the Graduierten\-kolleg ``Eichtheorien'' at the University of Mainz.


\setcounter{section}{0}
\setcounter{equation}{0}
\renewcommand{\theequation}{\mbox{\Alph{section}\arabic{equation}}}
\renewcommand{\thesection}{\Alph{section}}

\section{Appendix A}

 When integrating the spectrum functions it is convenient to transform to the
 integration variable $ \xi $, where $ x = y (\xi^2 + 1) / \xi $, such that 
 $ \beta =  (1 - \xi^2) / (1 + \xi^2) $.

 Trilog functions are generated from the integrals 

 \begin{Eqnarray} 
   \int_1^y \frac{1}{\xi} \ln (\xi) \ln(\xi - y) \, d \xi & = &
   \frac{\pi^2}{6} \ln y + \frac{1}{3} \ln^3 y -
   \Ti (y) + \zeta (3), \\[3mm]
   \int_1^y \frac{1}{\xi} \ln (\xi) \ln (1 - y \, \xi) \, d \xi & = & -
   \ln (y) \Li (y^2) - \Ti (y) + \Ti (y^2), \\[3mm]
   \int_1^y \frac{1}{\xi} \Li \Bigg( \frac{y (1 - \xi^2)}
    {\xi (1 - y \, \xi)} \Bigg) \, d \xi & = &
   \frac{1}{2} \ln (y) \Li (y^2) + 2 \,
   \Ti (y) - \Ti (y^2) - \zeta (3), \\[3mm]
   \int_1^y \frac{1}{\xi} \Li \Bigg( \frac{1 - \xi^2}
    {1 - y \, \xi} \Bigg) \, d \xi & = &
   \frac{\pi^2}{6} \ln y - \frac{1}{2} \ln (y) \Li (y^2) -
    2 \, \Ti (y) + \Ti (y^2) + \zeta (3), \quad
 \end{Eqnarray}
 
 where

 \begin{equation} 
   \Li (x) := - \int_0^1 \frac{\ln (1 - y)}{y} \, dy, \qquad
   \Ti (x) :=   \int_0^1 \frac{\Li (y)}{y} \, dy. 
 \end{equation}

 Use has been made of the relation

 \begin{equation} 
   \mbox{Li}_{n}(x) + \mbox{Li}_{n} (-x) =
   \frac{1}{2^{n-1}} \mbox{Li}_{n} (x^2).
 \end{equation}
 
 Euler's $ Zeta $ function is defined by
 
 \begin{equation} 
   \zeta (s) = \sum\limits_{k=1}^{\infty} k^{-s}, \qquad
   \zeta (3) = 1.202057 \ldots
 \end{equation}


\setcounter{equation}{0}

\section{Appendix B}

 The Fierz identity

 \begin{equation}
    \label{fierz1}
    [\gamma^{\mu} (1 - \gamma_5)]_{\alpha \beta} \,
    [\gamma_{\mu} (1 - \gamma_5)]_{\gamma \delta} = -
    [\gamma^{\mu} (1 - \gamma_5)]_{\gamma \beta} \,
    [\gamma_{\mu} (1 - \gamma_5)]_{\alpha \delta}
 \end{equation} 

 \noindent is well known.
 Not so well known is the Fierz identity (see e.g. \cite{scheck83}) 

 \begin{equation}
    \label{fierz2}
    [(1 \pm \gamma_5)]_{\alpha \beta} \,
    [(1 \mp \gamma_5)]_{\gamma \delta} = \frac{1}{2}
    [\gamma^{\mu} (1 \mp \gamma_5)]_{\gamma \beta} \,
    [\gamma_{\mu} (1 \pm \gamma_5)]_{\alpha \delta}.
 \end{equation} 

 \noindent Latter identity allows one to transform
 the $ q^\mu q^\nu $ piece of the $ W $--boson propagator
 discussed in Sec.~4 back into the standard form
 $ N^{\alpha \beta} C_{\alpha \beta} $ used in the
 remaining part of the paper. 


 \newpage

 \thispagestyle{empty}


 \newpage

 \begin{figure}[ht]

   \begin{picture}(150,165)
     \centerline{\psfig{figure=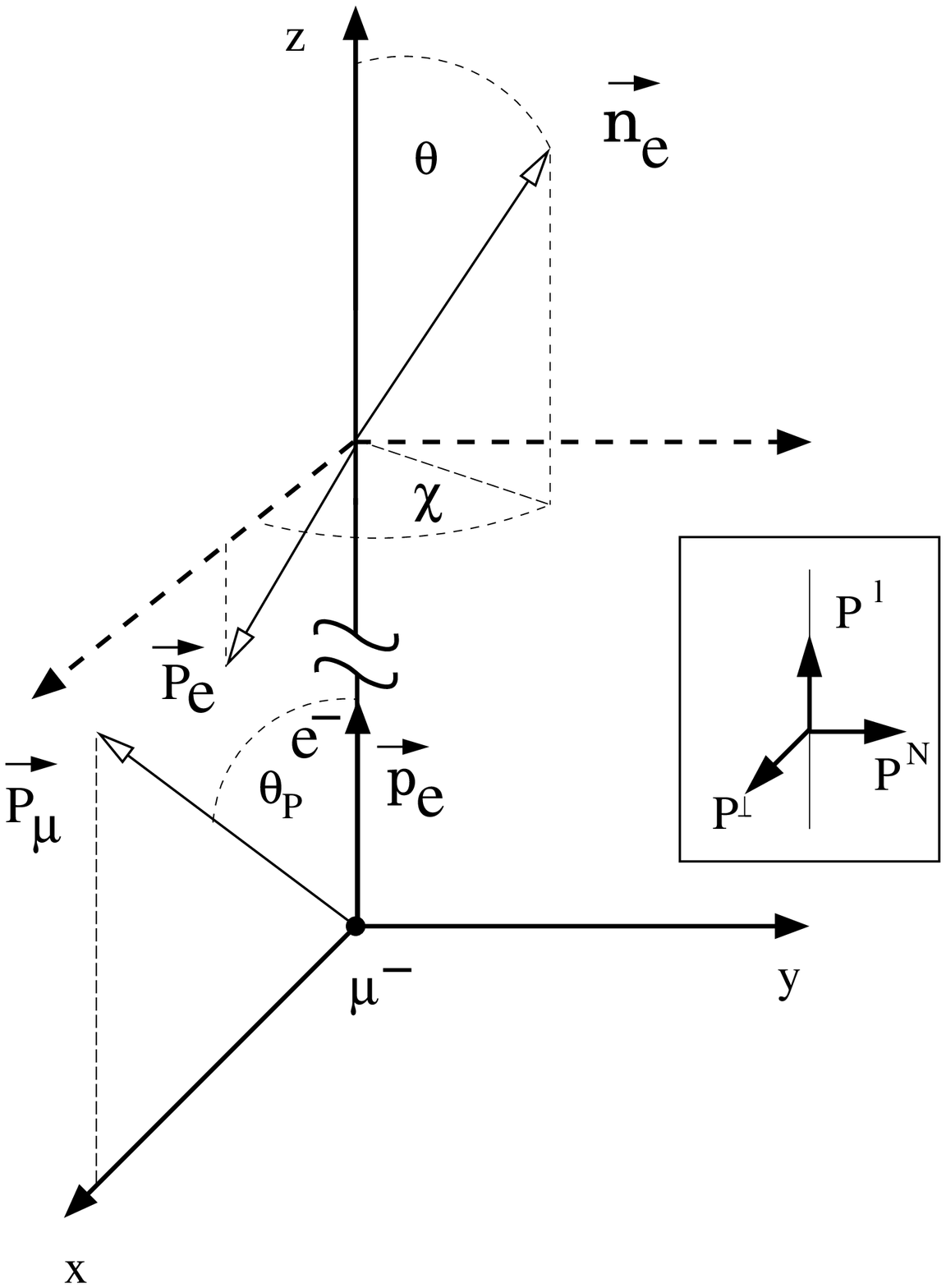, width=150mm}}
   \end{picture}

   \vspace*{2mm} \centerline{\bf Figure 1:} \vspace*{2mm}
   {\small Definition of the polar angles $ \theta $,
    $ \theta_P $ and the azimuthal angle $ \chi $.
    We have taken the artistic freedom to orient the polarization vector
    of the electron $ \vec{P}_{e} $ into the positive $ x $--direction
    contrary to what is calculated in the main text.}
 \end{figure}


 \newpage

 \begin{figure}[ht]

   \begin{picture}(155,092)
     \centerline{\psfig{figure=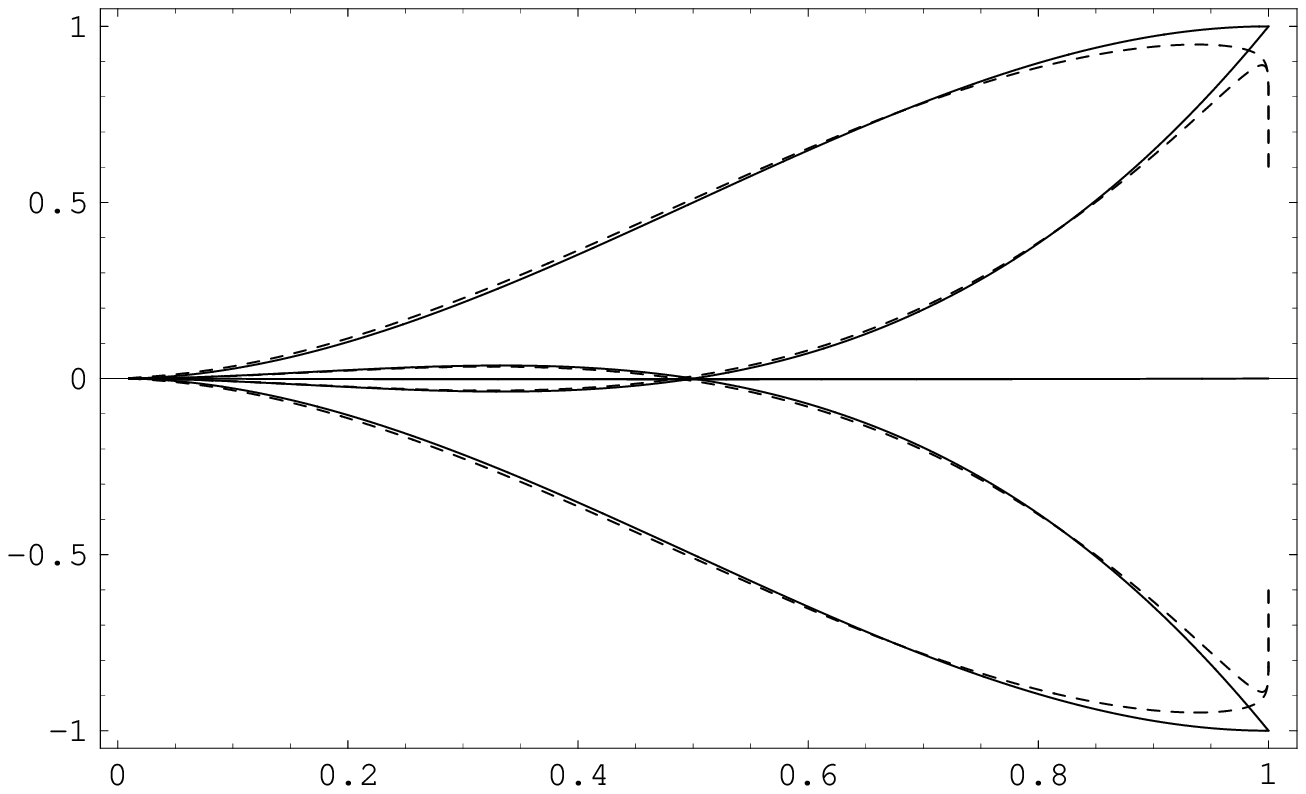, width=155mm, clip=}}
     \put(-070,-02){$ x $}
     \put(-160,+46){\rotatebox{90}{$ \beta \, x \, G_i $}}
     \put(-115,+85){--------- Born}
     \put(-115,+80){- - - - - Born + $ O(\alpha) $}
     \put(-125,+85){\bf 2a)}
     \put(-115,+20){\fbox{$ \mu \rightarrow e $}}
     \put(-55,+76){\small $ 1 $}
     \put(-55,+57){\small $ 4 $}
     \put(-55,+40){\small $ 2 $}
     \put(-55,+20){\small $ 3 $}
     \put(-55,+46){\small $ 5 $}
   \end{picture}

   \begin{picture}(155,100)
     \centerline{\psfig{figure=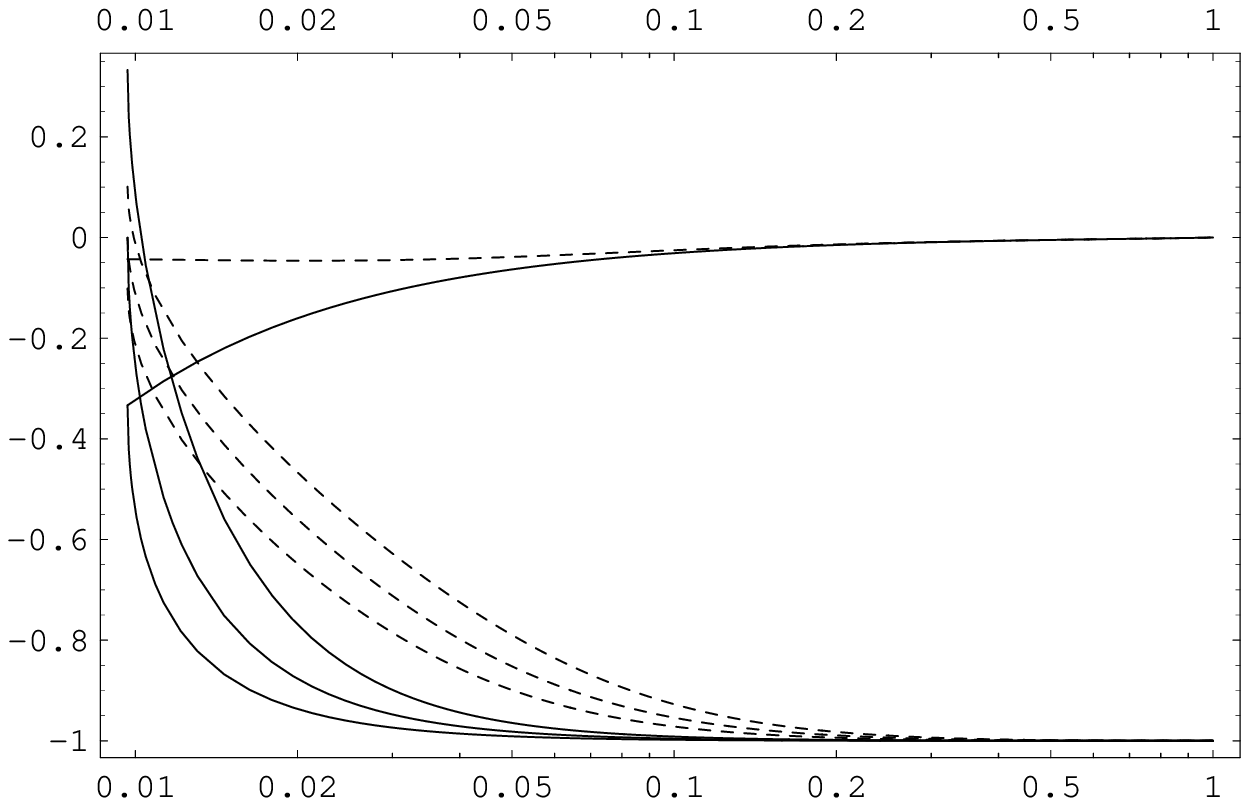, width=155mm, clip=}}
     \put(-070,-03){$ x $}
     \put(-160,+69){\rotatebox{90}{$ P_e^l, P_e^{\perp} $}}
     \put(-115,+85){--------- Born}
     \put(-115,+80){- - - - - Born + $ O(\alpha) $}
     \put(-125,+85){\bf 2b)}
     \put(-100,+35){$ P_e^l $}
     \put(-035,+60){$ P_e^{\perp} (\cos \theta = 0) $}
     \put(-135.5,+15.5){\tiny $ +1 $} \put(-121,+23){\tiny $ -1 $}
     \put(-103.0,+14.0){\tiny $ +1 $} \put(-094,+21){\tiny $ -1 $}
     \put(-126,+20){\tiny $  0 $} \put(-096.5,+18.5){\tiny $  0 $}
   \end{picture}
 \end{figure}

 \newpage

 \begin{figure}[th]
 
   \begin{picture}(155,105)
     \centerline{\psfig{figure=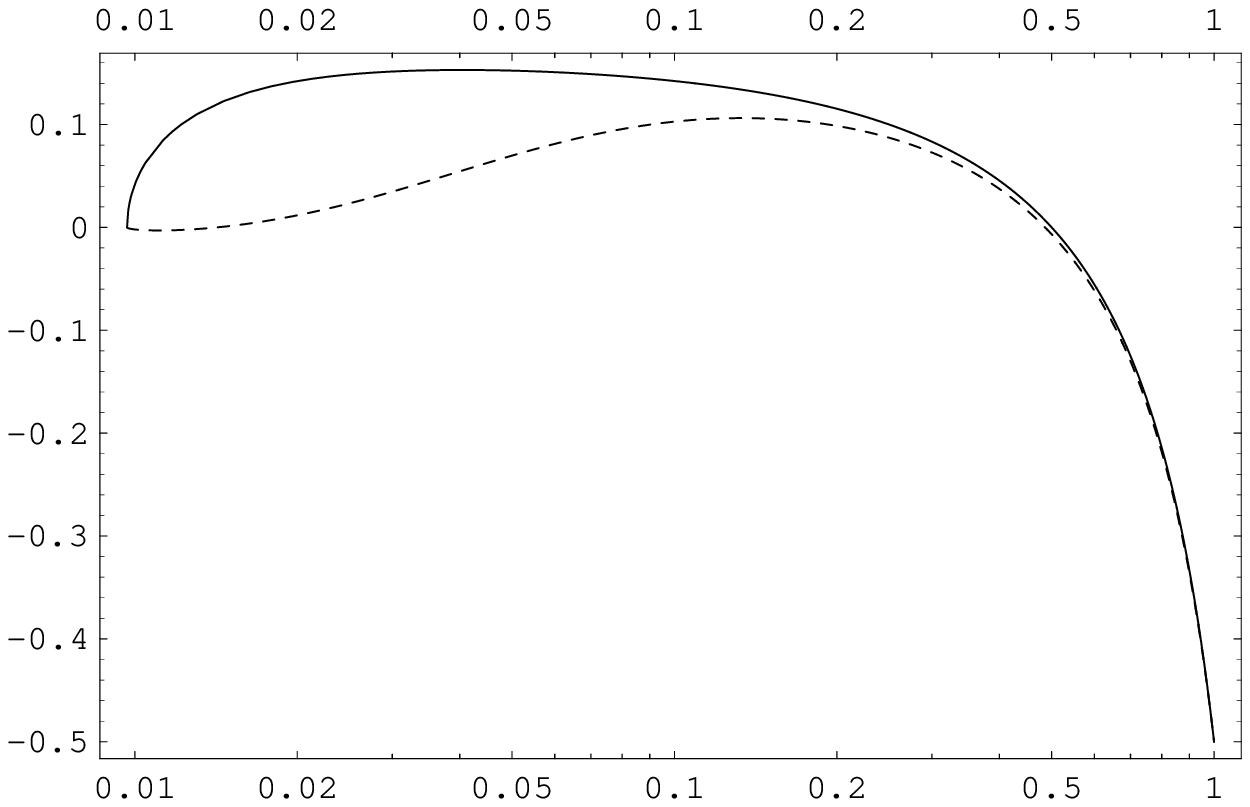, width=155mm, clip=}}
     \put(-070,-03){$ x $}
     \put(-160,+69){\rotatebox{90}{$ A_{FB} $}}
     \put(-115,+25){--------- Born}
     \put(-115,+20){- - - - - Born + $ O(\alpha) $}
     \put(-125,+25){\bf 2c)}
   \end{picture}
 
   \vspace*{10mm} \centerline{\bf Figures 2a, b, c:} \vspace*{2mm}
   {\small The case $ \mu \rightarrow e $. Scaled energy dependence of \newline
   a) spectrum functions $ \beta \, x \, G_i $, $ i = 1, \ldots, 5 $ \newline
   b) the longitudinal polarization of the electron $ P_e^l $
      for $ \cos \theta_P = -1, 0, +1 $, and \newline
   c) the forward--backward asymmetry $ A_{FB} $
      for $ \cos \theta = 0 $. \newline
      All curves with and without radiative corrections.
      Polarisation $ P $ is set to $ P = 1 $.}
 \end{figure}


 \newpage

 \begin{figure}[ht]

   \begin{picture}(155,092)
     \centerline{\psfig{figure=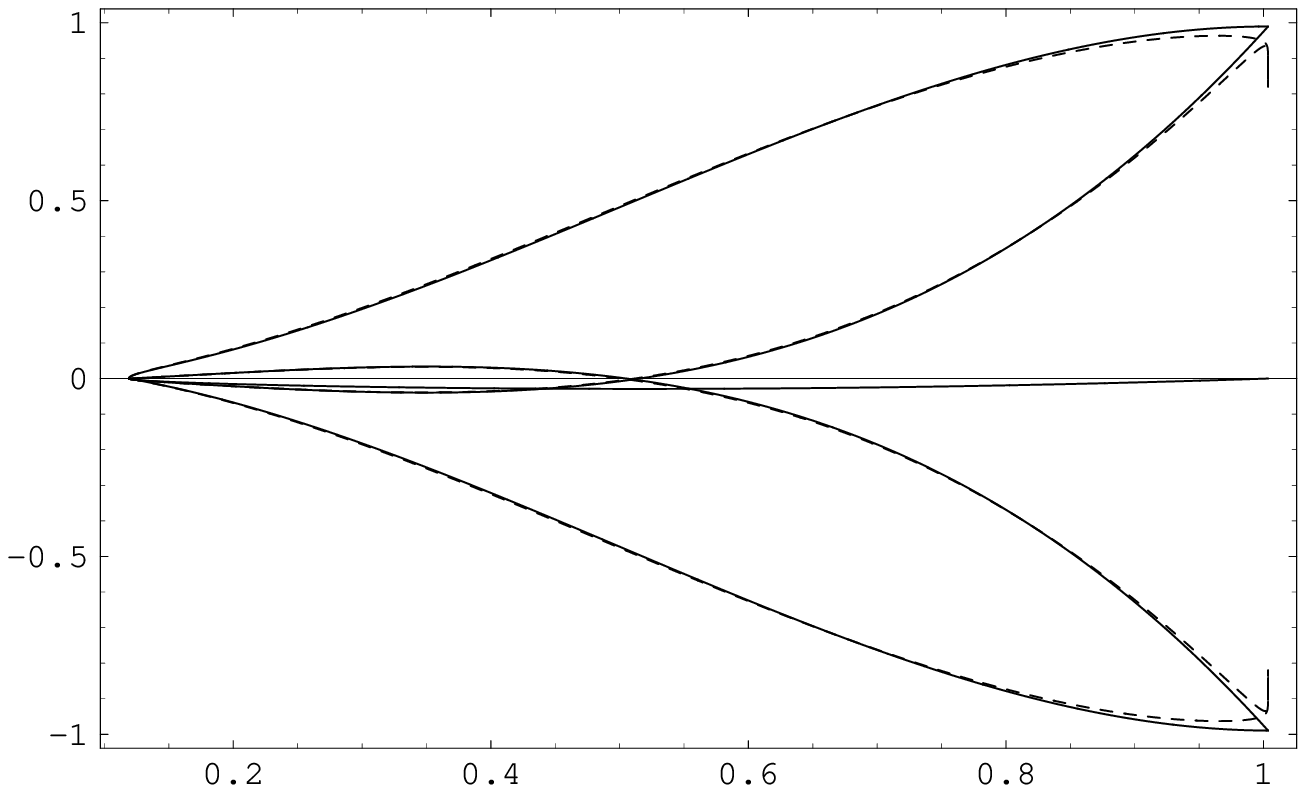, width=155mm, clip=}}
     \put(-070,-02){$ x $}
     \put(-160,+46){\rotatebox{90}{$ \beta \, x \, G_i $}}
     \put(-115,+85){- - - - - Born}
     \put(-115,+80){--------- Born + $ O(\alpha) $}
     \put(-125,+85){\bf 3a)}
     \put(-115,+20){\fbox{$ \tau \rightarrow \mu $}}
     \put(-55,+77){\small $ 1 $}
     \put(-55,+58){\small $ 4 $}
     \put(-55,+38){\small $ 2 $}
     \put(-55,+19){\small $ 3 $}
     \put(-55,+45){\small $ 5 $}
   \end{picture}

   \begin{picture}(155,100)
     \centerline{\psfig{figure=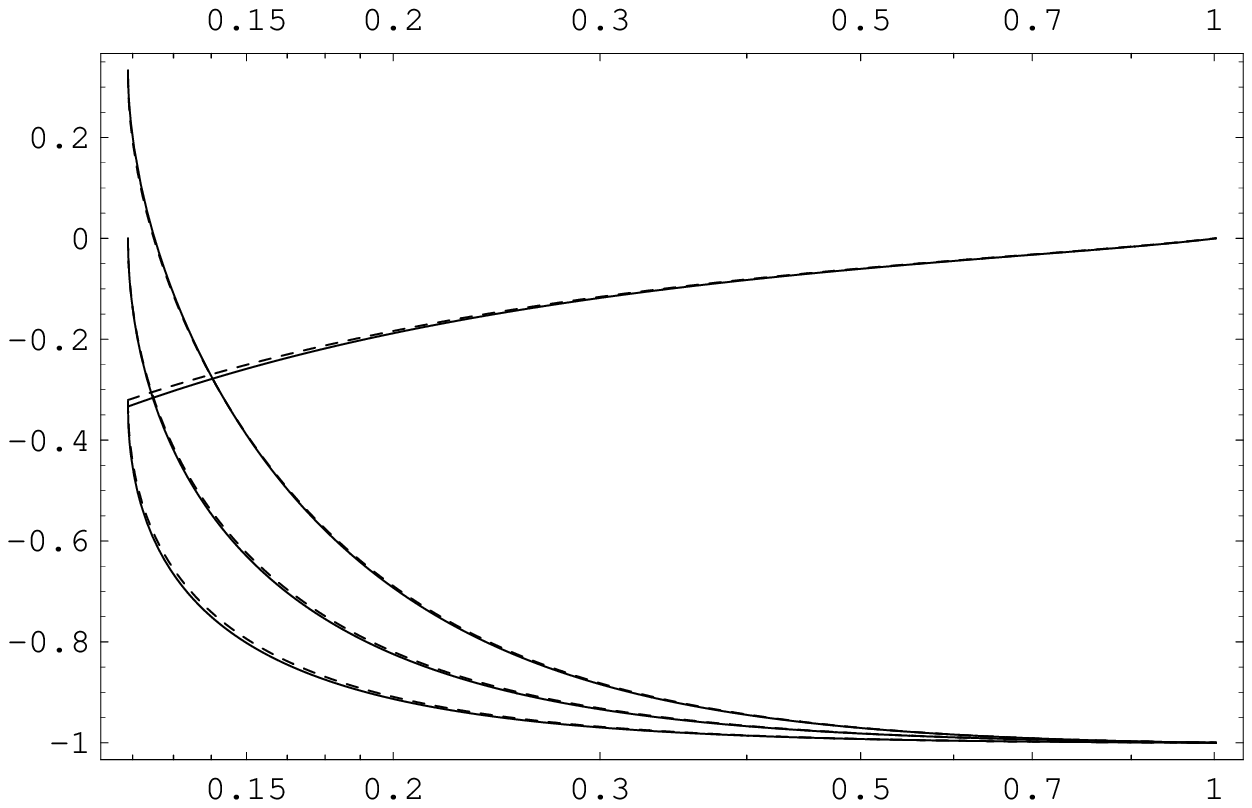, width=155mm, clip=}}
     \put(-070,-03){$ x $}
     \put(-160,+69){\rotatebox{90}{$ P_{\mu^l}, P_{\mu}^{\perp} $}}
     \put(-115,+85){- - - - - Born}
     \put(-115,+80){--------- Born + $ O(\alpha) $}
     \put(-125,+85){\bf 3b)}
     \put(-100,+35){$ P_{\mu}^l $}
     \put(-035,+60){$ P_{\mu}^{\perp} (\cos \theta = 0) $}
     \put(-124,+14){\tiny $ +1 $}
     \put(-112,+21){\tiny $  0 $}
     \put(-106,+26){\tiny $ -1 $}
   \end{picture}
 \end{figure}

 \newpage

 \begin{figure}[th]

   \begin{picture}(155,105)
     \centerline{\psfig{figure=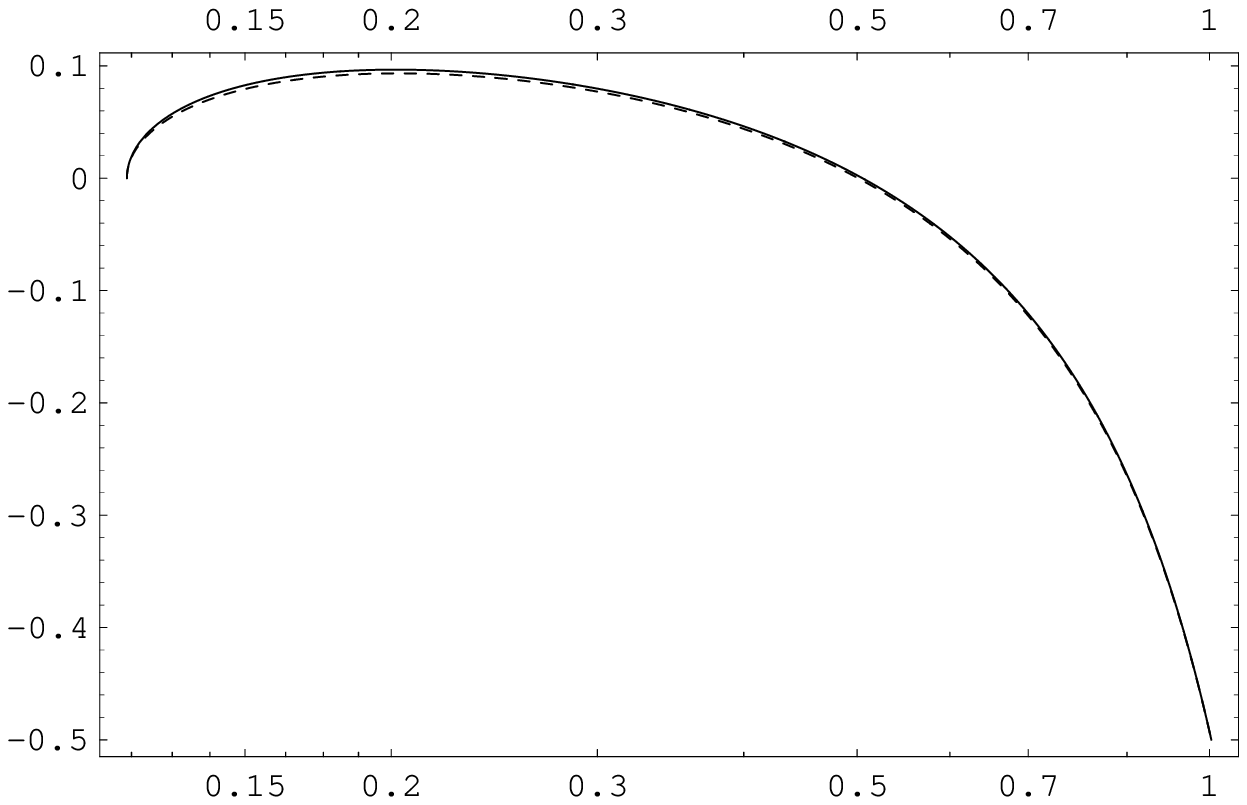, width=155mm, clip=}}
     \put(-070,-03){$ x $}
     \put(-160,+69){\rotatebox{90}{$ A_{FB} $}}
     \put(-115,+25){--------- Born}
     \put(-115,+20){- - - - - Born + $ O(\alpha) $}
     \put(-125,+25){\bf 3c)}
   \end{picture}
 
   \vspace*{10mm} \centerline{\bf Figures 3a, b, c:} \vspace*{2mm} 
   \centerline{\small The case $ \tau \rightarrow \mu $.
    Caption as in Fig.~2.}
 \end{figure}


 \newpage

 \begin{figure}[ht]

   \begin{picture}(155,092)
     \centerline{\psfig{figure=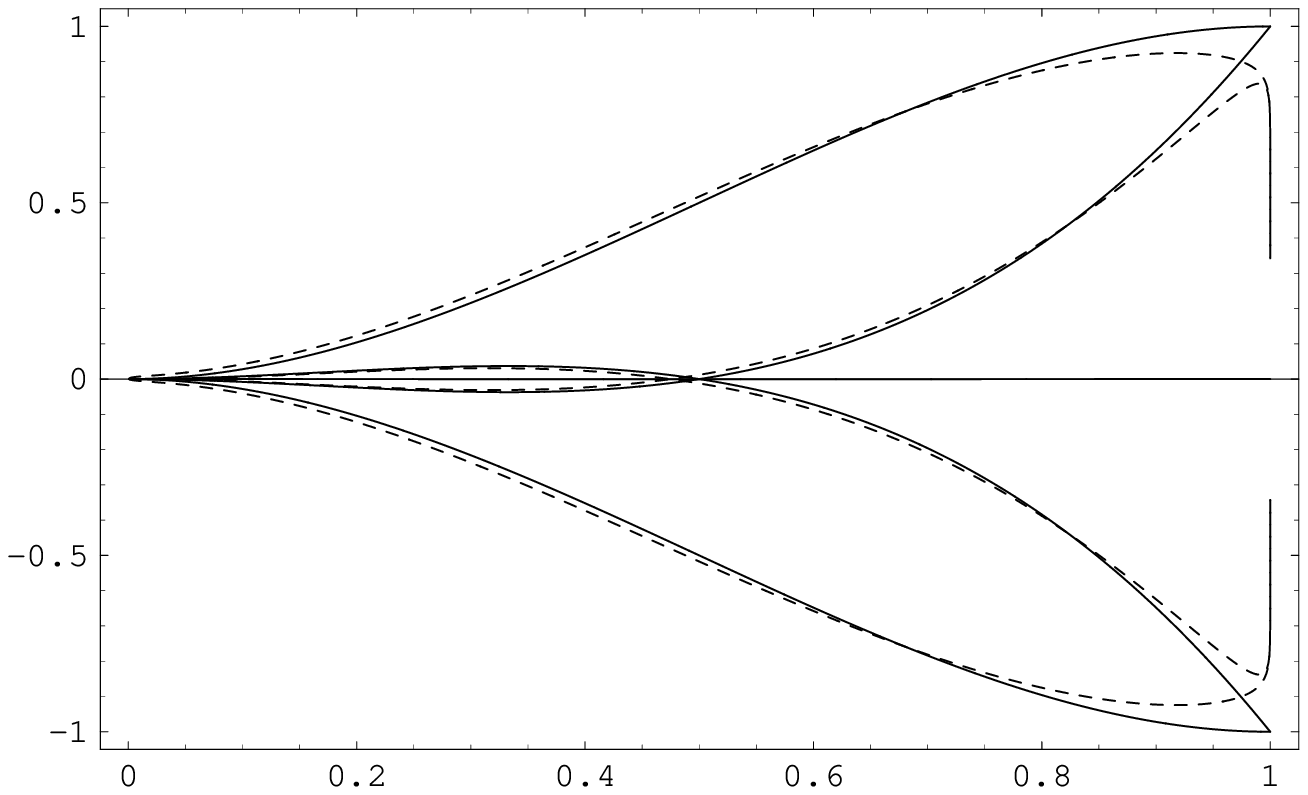, width=155mm, clip=}}
     \put(-070,-02){$ x $}
     \put(-160,+46){\rotatebox{90}{$ \beta \, x \, G_i $}}
     \put(-115,+85){--------- Born}
     \put(-115,+80){- - - - - Born + $ O(\alpha) $}
     \put(-125,+85){\bf 4a)}
     \put(-125,+20){\fbox{$ \tau \rightarrow e $}}
     \put(-55,+76){\small $ 1 $}
     \put(-55,+58){\small $ 4 $}
     \put(-55,+41){\small $ 2 $}
     \put(-55,+22){\small $ 3 $}
     \put(-55,+47){\small $ 5 $}
   \end{picture}

   \begin{picture}(155,100)
     \centerline{\psfig{figure=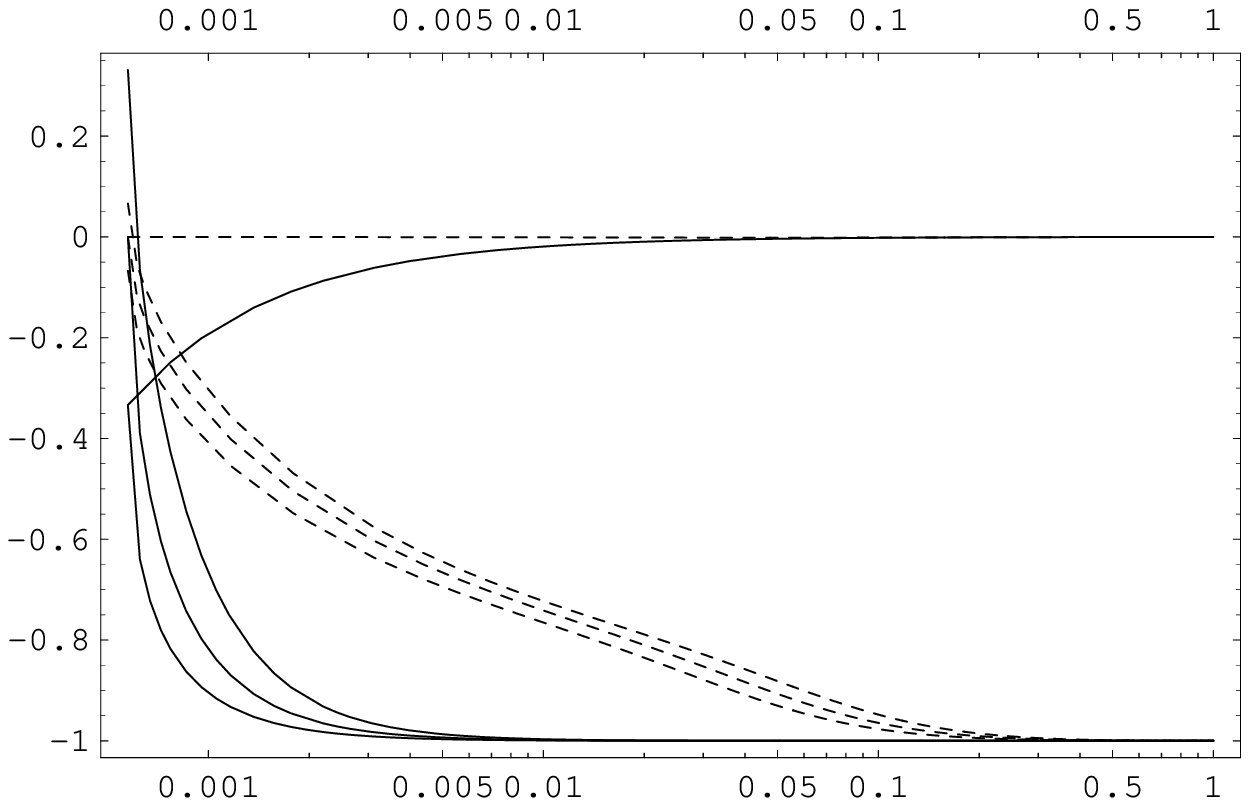, width=155mm, clip=}}
     \put(-070,-03){$ x $}
     \put(-160,+69){\rotatebox{90}{$ P_e^l, P_e^{\perp} $}}
     \put(-115,+85){--------- Born}
     \put(-115,+80){- - - - - Born + $ O(\alpha) $}
     \put(-125,+85){\bf 4b)}
     \put(-100,+35){$ P_e^l $}
     \put(-035,+60){$ P_e^{\perp} (\cos \theta = 0) $}
     \put(-137,+13){\tiny $ +1 $} \put(-125,+19.0){\tiny $ -1 $}
     \put(-095,+20){\tiny $ +1 $} \put(-087,+24.5){\tiny $ -1 $}
     \put(-129,+16){\tiny $  0 $}
   \end{picture}
 \end{figure}

 \newpage

 \begin{figure}[th]

   \begin{picture}(155,105)
     \centerline{\psfig{figure=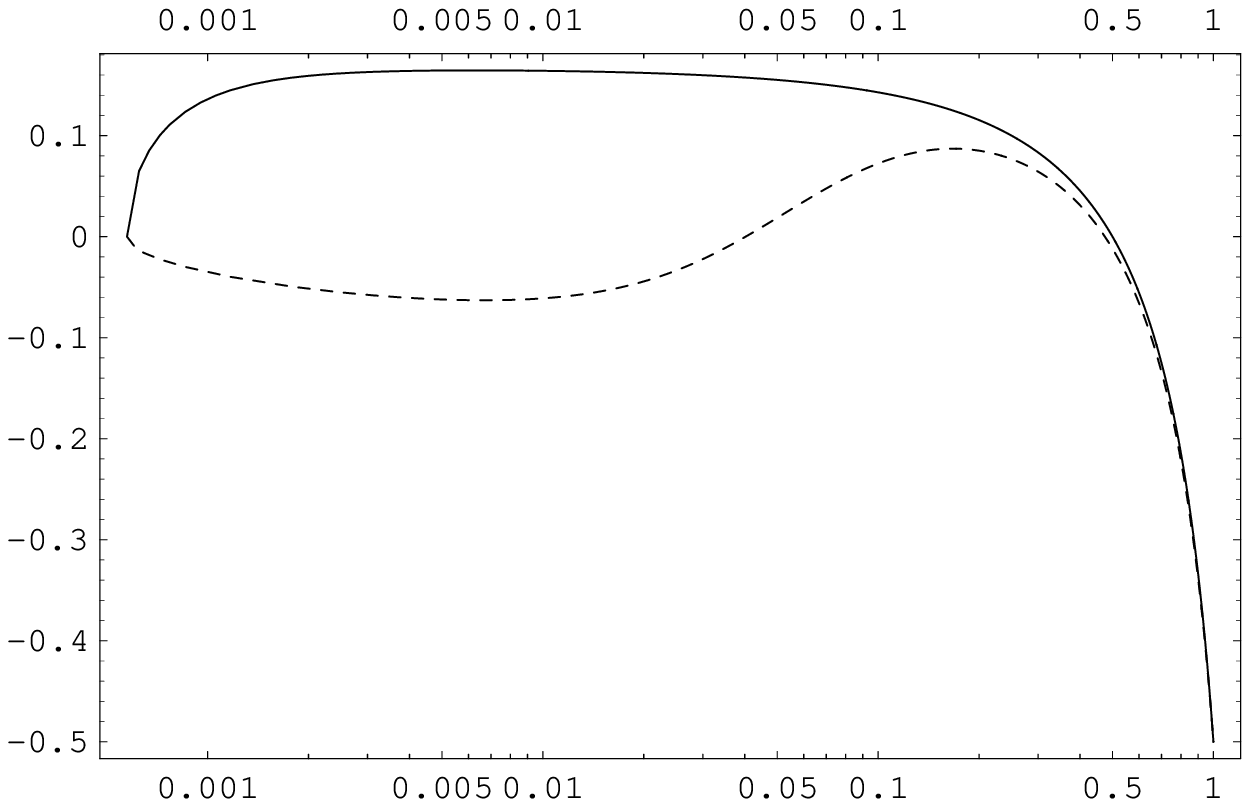, width=155mm, clip=}}
     \put(-070,-03){$ x $}
     \put(-160,+69){\rotatebox{90}{$ A_{FB} $}}
     \put(-115,+25){--------- Born}
     \put(-115,+20){- - - - - Born + $ O(\alpha) $}
     \put(-125,+25){\bf 4c)}
   \end{picture}
 
   \vspace*{10mm} \centerline{\bf Figures 4a, b, c:} \vspace*{2mm}
   \centerline{\small The case $ \tau \rightarrow e $.
    Caption as in Fig.~2.} 
 \end{figure}


 \newpage

 \begin{figure}[ht]

   \begin{picture}(160,105)
     \put(000,000){\psfig{figure=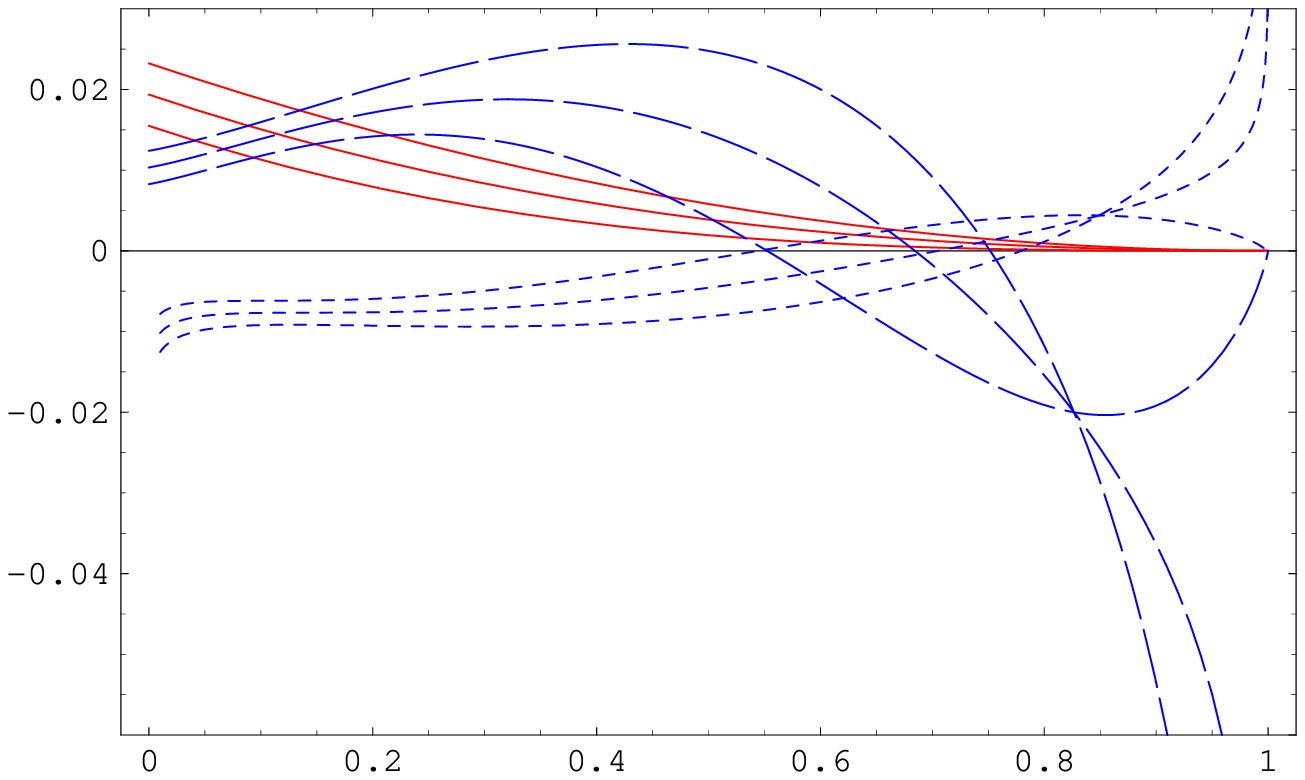, width=160mm, clip=}}
     \put(020,020){\fbox{$ \mu \rightarrow e $}}
     \put(030,090){\rot{$ 20 \times (h\!f) $}}
     \put(115,085){\bla{$ (n\!f; \ln y) $}}
     \put(040,045){\bla{$ (n\!f; \mbox{non-}\ln y) $}}
     \put(150,048){\bla{$ +1 $}}
     \put(150,010){\bla{$ \pp 0 $}}
     \put(135,010){\bla{$ -1 $}}
     \put(150,070){\bla{$ +1 $}}
     \put(150,077){\bla{$ \pp 0 $}}
     \put(135,077){\bla{$ -1 $}}
     \put(085,000){$ x $}
     \put(-05,030){\rotatebox{90}{$ d \Gamma^{h\!f/n\!f \, (\alpha)} / $
      $ (\Gamma_{\!0\,} dx \, d\!\cos \theta_P) $}}
   \end{picture}

   \vspace*{10mm} \centerline{\bf Figure 5:} \vspace*{2mm}
   {\small Scaled energy dependence of flip $ (h\!f) $ and
    no--flip $ (n\!f) $ spectrum functions 
    in the limit $ y \rightarrow 0 $.
    Latter contribution is separated into its non--logarithmic
    part $ (n\!f; \mbox{non--} \ln y) $ and its mass dependent
    logarithmic part $ (n\!f; \ln y) $ which is plotted for the
    case $ (\mu \rightarrow e) $.}
 \end{figure}

\end{document}